\documentclass[12pt]{article}
\usepackage[a0paper]{geometry}
\usepackage{anysize}
\marginsize{3cm}{3cm}{2cm}{2cm}
\usepackage{amsmath}
\usepackage{amsfonts}
\usepackage{bbm}
\usepackage{amssymb}
\usepackage{slashed}
\usepackage{mathrsfs}
\usepackage{graphicx}
\usepackage{epsfig}
\usepackage{enumerate}
\usepackage{dsfont}
\usepackage{appendix}
\usepackage{pdfpages}
\usepackage{setspace}
\usepackage{wrapfig}
\usepackage[utf8]{inputenc}
\usepackage{bm}
\usepackage{braket}
\usepackage{float}
\usepackage{epstopdf} 
\usepackage{authblk}
\usepackage{multicol}
\usepackage[
    backend=bibtex, 
    natbib=true,
    style=phys,
    giveninits=true,
    biblabel=brackets,
    abbreviate=false,
    doi=false, url=false, isbn=false,
    block=space,
    eprint=true
]{biblatex}
\newcommand{\e}{\text{e}}
\newcommand{\dG}{\dot{G}}
\newcommand{\ddG}{\ddot{G}}
\newcommand{\Li}{\text{Li}_2}
\newcommand{\dsc}{\text{Disc}}
\newcommand{\hs}{\hat{s}}
\newcommand{\hT}{\hat{t}}
\newcommand{\hu}{\hat{u}}

\title{Discontinuity method for evaluating scattering amplitudes within the worldline formalism}
\author{Victor Miguel Banda Guzm\'an \thanks{victor.banda@upslp.edu.mx} }
\affil{Universidad Polit\'ecnica de San Luis Potos\'i, Urbano Villal\'on 500, Col. La Ladrillera, C.P. 78363, San Luis Potos\'i, S.L.P., M\'exico.}
\date{}

\begin{document}

\maketitle

\begin{abstract}

    The worldline formalism offers an alternative framework to the standard diagrammatic approach in quantum field theory, grounded in first-quantized relativistic path integrals. Over recent decades, this formalism has attracted growing interest due to its potential applications and computational advantages. As a result, a collection of worldline master integrals has been derived within this approach. However, efficient mathematical tools for evaluating these integrals remain limited.

    Motivated by unitarity methods used in the conventional formalism of quantum field theory, this work proposes a novel framework for evaluating one-loop worldline master integrals, up to a rational function in the kinematical external invariants, through the computation of their discontinuities. Similarly to standard unitarity techniques, the method involves decomposing an $n$-point one loop amplitude in $D=4$ dimensions into a linear combination of tadpole, bubble, triangle, and box master integrals. The coefficients of this decomposition are rational functions of the external kinematic invariants. Each master integral exhibits a characteristic discontinuity with respect to a specific kinematic invariant. Consequently, the rational coefficients can be determined by directly computing the discontinuities of the corresponding worldline master integrals. In the three-point and four-point on-shell cases, this computation is particularly tractable, as the discontinuities involve Dirac delta functions arising from the Sokhotski–Plemelj formula, which simplify the resulting integrals.

    For the four-point on-shell case, a set of master formulas for the discontinuities is derived. The practical implementation of the proposed discontinuity method is demonstrated through several illustrative examples.
    
\end{abstract}

\section{Introduction}

In recent years, the worldline formalism has emerged as a compelling alternative to the conventional diagrammatic techniques employed in perturbative quantum field theory. Originating from Feynman’s work in the early 1950s on quantum electrodynamics (QED) \cite{FeynmanQED,FeynmanOpCal}, this approach is based in first-quantized relativistic path integrals. Since its introduction, considerable research has been devoted to expanding the formalism’s range of applicability. Among its many uses, it has been employed to:

\begin{enumerate} 
    \item Obtain effective actions \cite{Wl_effective_act1,Wl_effective_act2}.
    \item Determine quantum anomalies \cite{WL_anomalies}.
    \item Compute multi-loop amplitudes \cite{MultiloopAmp}.
    \item Describe quantum processes in electromagnetic background fields \cite{WL_Shaisultanov,WL_Adler,WL_Reuter,ExternalField5,ExternalField6,ExternalField7,WL_Ilderton,plane_wave_James,ExternalField3,ExternalField4,ExternalField2,ExternalField1,ExternalField8}.
    \item Describe axial and Yukawa couplings \cite{AY2,WL_Hoher1, WL_Hoher2, AY1, AY3}.
    \item Compute $N$-gluon amplitudes \cite{WL_QCD1,WL_QCD2,WL_QCD3}.
    \item Compute QED quantum amplitudes in vacuum \cite{WL_QED4,WL_QED2,WL_QED3,WL_QED1}.
    \item Describe quantum processes with gravitational interactions \cite{WL_grav1, WL_grav2}.    
\end{enumerate}

The examples mentioned above do not constitute an exhaustive account of the developments achieved over the past decades within the worldline formalism. For a more detailed review and further references, see Ref. \cite{Schubert_WL}. 

Despite the significant advances made within the worldline formalism, several challenges remain to be addressed. Among them is the ongoing need for mathematical tools that facilitate the evaluation of the integrals arising from the formalism. Consider for example the case of scalar QED. The one-loop $N$-photon amplitude $\Gamma_{\text{scal}}$ in $D$-dimensions from the Worldline formalism reads as \cite{Polyakov_book,Strassler,WL_QED4,Schubert_WL,WL_QED3,}, 
\begin{eqnarray}
    \Gamma_{\text{scal}}(p_1, \epsilon_1; \dots ; p_N, \epsilon_N) &=& (-ie)^N (2\pi)^D \delta^D\left( \sum k_i \right) \int_0^\infty \dfrac{dT}{T} (4\pi T)^{\frac{-D}{2}} \e^{-m^2T} \nonumber \\
    && \hspace{-120pt} \times  \int_0^T \prod_{k=1}^N d\tau_k \, \text{exp}\left.\left\{ \sum_{i,j=1}^N \left[ \dfrac{1}{2} G_{ij} p_i \cdot p_j - i \dG_{ij} \epsilon_i \cdot p_j + \dfrac{1}{2} \ddG_{ij} \epsilon_i \cdot \epsilon_j  \right] \right\}\right\vert_{\epsilon_1 \dots \epsilon_N}, \label{wl_sqed_formula}
\end{eqnarray}
where $m$ and $e$ corresponds to the mass and electric charge of the electron, $p_i$ and $\epsilon_i$ represents the momenta and vector polarization of the external photons, and the abbreviation in the Worldline Green's function $G_{ij}$ stands for $G(\tau_i,\tau_j)$, which are defined as, 
\begin{eqnarray}
    G(\tau_i,\tau_j) &=& |\tau_i-\tau_j| - \dfrac{(\tau_i-\tau_j)^2}{T}, \nonumber \\
    \dG(\tau_i,\tau_j) &=& \text{sign}(\tau_i-\tau_j) - 2 \dfrac{\tau_i-\tau_j}{T}, \nonumber \\
    \ddG(\tau_i,\tau_j) &=& 2 \delta(\tau_i-\tau_j) - \dfrac{2}{T}. \label{def_Gs_wl}
\end{eqnarray}
Here, a dot denotes a derivative with respect to the first variable. The notation $\vert_{\epsilon_1 \dots \epsilon_N}$ means that the exponential should be expanded, and only the terms linear in each of the polarization vectors be kept. 

Eq. \ref{wl_sqed_formula} collects into a single expression the complete set of Feynman diagrams corresponding to the one-loop interaction of $N$ photons. However, its direct evaluation requires a non negligible effort even when all the external momenta are on-shell. This highlights the need for alternative approaches to efficiently address the integrals that arise in worldline formulations.

On the other hand, in the standard diagrammatic approach of quantum field theory several tools for evaluating Feynman integrals have been developed in the last decades. We have for example, the integration by parts technique \cite{IBP_alg, IBP_letter}, tensor reduction methods \cite{TensorRed_Dav,TensorRed_Anastasiou,TensorRed_Fior}, dimensional shift relations \cite{DimensionalShift_Tarasov1, DimensionalShift_Tarasov2}, the method of differential equations \cite{DifEq_Kotikov1, DifEq_Kotikov2, DifEq_Remiddi1, DifEq_Remiddi2}, and in particular for one loop integrals, the Passarino Veltman reduction \cite{PV_Red,PV_Red_Pittau1,PV_Red_Pittau2,PV_Red_Pittau3,PV_Red_Pittau4,PV_Red_Weinzierl,PV_Red_Weinzierl2} and unitarity based methods \cite{UnitarityMeth_Bern1, UnitarityMeth_Bern2, UnitarityMeth_Britto1}. 

In unitarity-based methods, a central feature is the observation that any one-loop amplitude can be expressed as a linear combination of a finite set of master integrals, with coefficients that are rational functions of the internal masses and external kinematic invariants \cite{UnitarityMeth_Britto1}. 

\begin{figure}
    \centering
    \includegraphics[width=0.4\linewidth]{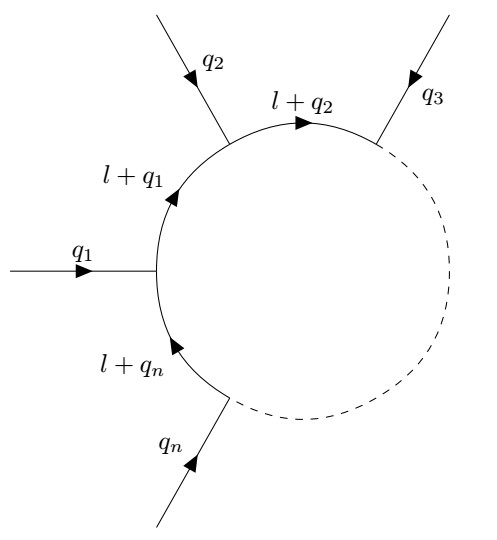}
    \caption{Scalar Feynman diagram with $n$-external legs. }
    \label{npointdiagram}
\end{figure}

Thus, for $D=4$ dimensions, a one-loop amplitude $A^{(1)}$ in the dimensional regularization scheme decomposes as \cite{Book_Weinzierl},
\begin{eqnarray}
    A^{(1)} &=& \sum_{i_1<i_2<i_3<i_4} c_{i_1i_2i_3i_4} I_4(i_1,i_2,i_3,i_4) + \sum_{i_1<i_2<i_3} c_{i_1i_2i_3} I_3(i_1,i_2,i_3) \nonumber \\
    && + \sum_{i_1<i_2} c_{i_1i_2} I_2(i_1,i_2) + \sum_{i_1} c_{i_1} I_1(i_1) + R + \mathcal{O}(\epsilon), \label{General_oneloop_amp}
\end{eqnarray}
where $\epsilon$ is the dimensional regularization parameter, and the functions $I_n(i_1, \dots, i_n)$ corresponds to the scalar $n$-points integral functions,
\begin{eqnarray}
    I_n(i_1, \dots, i_n) &=& e^{\epsilon \gamma_E} \mu^{2 \epsilon} \int \dfrac{d^D l}{i \pi^{\frac{D}{2}}} \dfrac{1}{d_{i_1} \dots d_{i_n}}, \label{npoint_functions}
\end{eqnarray}
which are expanded up to order $\mathcal{O}(\epsilon)$. The denominators $d_{i}$ of the equation \eqref{npoint_functions} are defined as \footnote{The conventions adopted in this manuscript are detailed in Appendix \ref{Conventions}.}, 
\begin{eqnarray}
    d_{i} = (l+q_i)^2+m_i^2.
\end{eqnarray}
Here, $m_i$ and $q_i$ corresponds to the internal masses and momenta of the external legs of the scalar Feynman diagram depicted in Fig. \ref{npointdiagram}, respectively. 

The coefficients $c_{i_1 \dots i_n}$ and the rational term $R$ in the Eq. \ref{General_oneloop_amp} are rational functions in the internal masses and external kinematical invariants, and are independent of the dimensional regularization parameter. Since each master integral $I_n$ has its own characteristic discontinuity across a branch cut for a given kinematical region, we can reconstruct the one-loop amplitude up to a rational term from unitarity cuts \cite{UnitarityMeth_Britto1}. Thus, the terms in Eq. \ref{General_oneloop_amp} that depends on the $n$-point integral functions $I_n$ are called cut constructable. 

In this work, we adapt concepts from unitarity-based methods to facilitate the evaluation of worldline master integrals, such as the one presented in Eq. \ref{wl_sqed_formula}. Specifically, we compute the discontinuity of the four-point on-shell worldline integrals in a given kinematic region and match it to the discontinuity of a corresponding master Feynman integral. This procedure allows us to extract the cut-constructible part of the worldline integral. 

The article is organized as follows: Sec. \ref{Worldline formalism} contains an introduction to the worldline formalism for $N$-point one-loop amplitudes in massive $\phi^3$ field theory and scalar QED. It reviews the derivation of the worline master integrals for these theories. In Sec. \ref{One-mass_triangle_section} the discontinuity of one-mass triangle integrals within the worldline formalism for $D=4$ dimensions is derived. It introduces the main integration techniques that will be used in the rest of the article. In sec. \ref{Four_point_wl_integrals} the discontinuity of the four-point on-shell one-loop worldline integrals are derived. The discontinuities are evaluated with respect to the $s$ Mandelstam variable, in $D=4$ dimensions. A set of master formulas for these discontinuities is presented at the end of the section. Sec. \ref{Disc_method_examples} outlines the application of previously derived master formulas for discontinuities to the evaluation of worldline integrals. In particular, it focuses on the four-point scalar amplitude in $\phi^3$ field theory and one of the principal worldline integrals contributing to the four-photon scattering amplitude in scalar QED. The concluding remarks and a summary of the work are provided in Sec. \ref{Conclusion_summary}. The manuscript is complemented with five appendixes which contains supplemental information. Appendix \ref{Conventions} lists the conventions of the manuscript. Appendix \ref{Integration_path_integrals} provides a derivation for the Gaussian integration of the worldline scalar path integrals. Appendix \ref{DiracDelta_iden_Box} provides a detailed derivation of the principal integral identity employed in Sec. \eqref{Four_point_wl_integrals} for the evaluation of discontinuities. Appendix \ref{Disc_Box_Append} presents the computation of the discontinuities of the function $D(x,y)$, which is proportional to the on-shell box integral and is defined in Eq. \eqref{Dav_fun}, using the coproduct formalism for multiple polylogarithms. Appendix \ref{Two_point_wl_int} contains the evaluation of two-point amplitudes within the worldline formalism in the dimensional regularization scheme.            

\section{Worldline master formulas for $N$-point one-loop amplitudes in $\phi^3$ field theory and scalar QED} \label{Worldline formalism} 

This section presents a review of the master formulas for $N$-point one-loop amplitudes in $\phi^3$ field theory and scalar QED. Although these theories are relatively simple, they are employed as illustrative examples to highlight the general characteristics of the integrals that emerge in the worldline formalism.

In the simplest case of $\phi^3$ field theory, the one-loop effective action $\Gamma[\phi]$ within the worldline formalism possesses the following path integral representation in Euclidean spacetime \cite{Schubert_WL},
\begin{eqnarray}
    \Gamma[\phi] = \dfrac{1}{2} \int_0^\infty \dfrac{dT}{T} e^{-m^2T} \int_{x(T)=x(0)} Dx\, \e^{-\int_0^T d\tau \left[ \frac{1}{4}\dot{x}^2(\tau) + \lambda \phi( x(\tau) ) \right]  }. \label{Effective_phi3}
\end{eqnarray}
Here the path integral is performed over all closed trajectories in spacetime with periodicity $T$, with the parameter $T$ representing the proper time of the scalar particle in the loop. 

The $N$-point one loop amplitudes for the $\phi^3$ field theory can be obtained from the effective action given in the equation \eqref{Effective_phi3} by specializing the background field $\phi$ to a sum of plane waves,
\begin{eqnarray}
    \phi(x) = \sum_{i=1}^N \e^{i p_i x}.
\end{eqnarray}
Substituting this expansion in the equation \eqref{Effective_phi3}, the exponential term with the background field $\phi$ can be rewritten as, 
\begin{eqnarray}
    \e^{- \lambda \int_0^T d\tau\, \phi} = \sum_{k=0}^\infty \dfrac{(-\lambda)^k}{k!} \left[ \sum_{i=1}^N \int_0^T d\tau\, \e^{i p_i x(\tau)}\right]^k, \label{phi_expansion}
\end{eqnarray}  
By picking out the terms containing every $p_i$ precisely once (which also removes the $1/k!$ factor in the equation \eqref{phi_expansion} ), the $N$-point one loop amplitude $\Gamma_{\text{1PI}}$ that arise from one-particle irreducible Feynman diagrams reads as,
\begin{eqnarray}
    \Gamma_{\text{1PI}}(p_1,\dots,p_N) &=& \dfrac{1}{2} (-\lambda)^N \int_0^\infty \dfrac{dT}{T} e^{-m^2T} \int_{x(T)=x(0)} Dx\, \left( \prod_{i=1}^N  \int_0^T d\tau_i\, \e^{i p_i \cdot x(\tau_i) } \right) \nonumber \\
    && \times \e^{- \frac{1}{4} \int_0^T d\tau\, \dot{x}^2(\tau)  }.
    \label{scalar_Npoint}
\end{eqnarray}

To perform the path integral in the equation \eqref{scalar_Npoint}, one can split the closed trajectories $x(\tau)$ with periodicity $T$ as,
\begin{eqnarray}
    x(\tau) = q(\tau) + x_0, \label{split_trajec}
\end{eqnarray}
where $x_0 = x(0)$. 

From equation \eqref{split_trajec}, $q(\tau)$ satisfies the Dirichlet boundary conditions, 
\begin{eqnarray}
    q(0) = q(T) = 0. \label{BC_q}
\end{eqnarray}

The decomposition presented in equation \eqref{split_trajec} is not unique, as alternative boundary conditions for $q(\tau)$ and different definitions of $x_0$ can be adopted. For example, there is the string-inspired convention, \cite{Schubert_WL,WL_notes}, where $x_0$ corresponds to the center-of-mass (or average position) of the loop,
\begin{eqnarray}
    x_0 = \dfrac{1}{T} \int_0^T d\tau\, x(\tau).
\end{eqnarray}
Hence, $q(\tau)$ in addition to periodicity, satisfies the non-local boundary condition,
\begin{eqnarray}
    \int_0^T d\tau \, q(\tau) = 0. \label{non_local_bc_q}
\end{eqnarray}

Splitting the worldline trajectories $x(\tau)$ according to the equation \eqref{split_trajec}, and from the following identities
\begin{eqnarray}
    \int Dx = \int dx_0^D \int Dq, \qquad \int dx_0^D\, \e^{i x_0 \cdot \sum_i p_i} = (2\pi)^D \delta(\sum_i p_i),
\end{eqnarray}
the expression in equation \eqref{scalar_Npoint} rewrites as,
\begin{eqnarray} 
    \Gamma_{\text{1PI}}(p_1,\dots,p_N) &=& \dfrac{1}{2} (-\lambda)^N (2\pi)^D \delta(\sum_{i=1}^N p_i) \int_0^\infty \dfrac{dT}{T} e^{-m^2T} \nonumber \\
    && \times \int_{\text{DBC}} Dq\, \int_0^T \prod_{k=1}^N d\tau_k \, \e^{ \sum_{i=1}^N i p_i \cdot q(\tau_i) - \frac{1}{4} \int_0^T d\tau\, \dot{q}^2(\tau)  },
    \label{scalar_Npoint_v2} 
\end{eqnarray}
where the path integral $\int_{\text{DBC}}$ is taken over all functions $q(\tau)$ that satisfies the Dirichlet boundary conditions. 

The evaluation of the path integral in equation \eqref{scalar_Npoint_v2} can be performed using Gaussian integration, as shown in the Appendix \ref{Integration_path_integrals}. Using the identity \eqref{path_integral_scalar2} of this appendix and conservation of momentum, the $N$-point one loop amplitude $\Gamma_{\text{1PI}}$ in the equation \eqref{scalar_Npoint_v2} reduces to,
\begin{eqnarray}
    \Gamma_{\text{1PI}}(p_1,\dots,p_N) &=& \dfrac{1}{2} (-\lambda)^N (2\pi)^D \delta(\sum_{i=1}^N p_i) \int_0^\infty \dfrac{dT}{T} e^{-m^2T} \left(4\pi\, T\right)^{-\frac{D}{2}} \nonumber \\
    && \times \int_0^T \prod_{k=1}^N d\tau_k\,\e^{\frac{1}{2} \sum_{i,j=1}^N p_i \cdot p_j\, G_{i,j} }. \label{scalar_phi3_masterformula}
\end{eqnarray}

For scalar QED, the path integral representation of the one-loop effective action $\Gamma_{\text{scal}}[A]$ reads as \cite{Schubert_WL,WL_notes}, 
\begin{eqnarray}
    \Gamma_{\text{scal}}[A] = \int_0^\infty \dfrac{dT}{T} e^{-m^2T} \int_{x(T)=x(0)}  Dx \, \e^{-\int_0^T d\tau\, \left[ \frac{1}{4} \dot{x}^2 + ie\dot{x}^\mu A_\mu(x) \right] },
\end{eqnarray}
where $A_\mu(x)$ corresponds to the vector potential of the electromagnetic field. Compared with equation \eqref{Effective_phi3}, note that now there is no global factor of $\frac{1}{2}$ since the scalar field coupled to the electromagnetic field has two real degrees of freedom.  

Specializing the potential $A_\mu(x)$ to a sum of planes waves with definite momenta and polarizations,
\begin{eqnarray}
    A^\mu(x) = \sum_{i=1}^N \epsilon_i^\mu \, \e^{i p_i \cdot x}
\end{eqnarray}
one can proceed analogously as in the case of the $\phi^3$ field theory, and obtain the master formula given in the equation \eqref{wl_sqed_formula} for the the one-loop $N$-photon amplitude $\Gamma_{\text{scal}}$.

By performing the integration over $T$ with a rescaling of the worldline parameters as $\tau_i = T u_i$, the remaining integrals in equations \eqref{wl_sqed_formula} and \eqref{scalar_phi3_masterformula} take the following form,
\begin{equation}
    \int_0^1 \prod_{k=1}^N du_k\, \dfrac{P( \dG, \ddG )}{\left[m^2-\frac{1}{2} \sum_{i,j=1}^N p_i \cdot p_j\, G_{ij}\right]^M}, \label{wl_integrals}
\end{equation}
where $P$ is a polynomial function on $\dG$ and $\dG$, and the Green's function $G_{ij} := G(u_i,u_j)$, and its derivatives are defined now as,
\begin{eqnarray}
    G(u,u^\prime) &=& |u-u^\prime| - (u-u^\prime)^2, \nonumber \\
    \dG(u,u^\prime) &=& \text{sign}(u-u^\prime)-2(u-u^\prime), \nonumber \\
    \ddG(u,u^\prime) &=& 2 \delta(u-u^\prime)-2. \label{G_def_scaled}
\end{eqnarray}

For spinor quantum electrodynamics (QED) and quantum chromodynamics (QCD), the worldline principal integrals associated with the master formulas for one-loop $N$-point amplitudes exhibit a structure similar to that of equation \eqref{wl_integrals}, \cite{Schubert_WL}. 

To evaluate the master integrals arising from the worldline formalism, this work proposes an approach based on the discontinuities of the integrals, analogous to the method employed in the standard formalism. This approach is illustrated through the one-loop four-photon amplitude, with on-shell photons, in scalar QED in four dimensions. As a preliminary introduction to the techniques of the discontinuity method, in the next section the discontinuities of the 3-point amplitude $\Gamma_{\text{1PI}}(p_1,p_2,p_3)$ where $p_2^2=p_3^2=0$, often referred to in the literature as one-mass triangle integrals (see for example \cite{Britto_triangle_diagrams}), is computed.

\section{Discontinuity of one-mass triangle integrals within the world line formalism} \label{One-mass_triangle_section}

Consider the case with $N=3$ in equation \eqref{scalar_phi3_masterformula} in $D=4$ dimensions. After integration overt $T$, and rescaling the worldline parameters, the scalar amplitude $\Gamma_{\text{1PI}}$ is proportional to the integral $T(q_1,q_2,q_3)$ defined as,
\begin{eqnarray}
    T(q_1,q_2,q_3) &=& \int_0^1 \prod_{k=1}^3 du_k \, \dfrac{1}{m^2 \left( 1 + q_1 \Lambda_1 + q_2 \Lambda_2 + q_3 \Lambda_3 \right) }, \label{Triangle_integral}
\end{eqnarray}
where $q_i=\frac{p_i^2}{m^2}$ and
\begin{eqnarray}
    \Lambda_1 &=& \dfrac{1}{2} \left( G_{12} + G_{13} -G_{23} \right), \nonumber \\
    \Lambda_2 &=& \dfrac{1}{2} \left( G_{21} + G_{23} -G_{13} \right), \nonumber \\
    \Lambda_3 &=& \dfrac{1}{2} \left( G_{31} + G_{32} -G_{12} \right), \label{Def_Lambdas}
\end{eqnarray}
with the Green's functions $G_{ij}$ defined in the equation \eqref{G_def_scaled}.

Setting $q_2=q_3=0$ (the one-mass triangle case), the integral $T(q_1,q_2,q_3)$ in the equation \eqref{Triangle_integral} reduces to,
\begin{eqnarray}
    T(q_1,0,0) &=& \int_0^1 \prod_{k=1}^3 du_k \, \dfrac{1}{m^2 \left( 1 + q_1 \Lambda_1 \right) }. \label{Triangle_integral_one_mass}
\end{eqnarray}
By translational invariance on the parameters $u_i$, one can set for example $u_3=0$, and split the remaining integrals into two sectors, namely: $0 \leq u_2 \leq u_1 \leq 1$ and $0 \leq u_1 \leq u_2 \leq 1$. Thus,
\begin{eqnarray}
    T(q_1,0,0) &=& \dfrac{1}{m^2} \Bigg( \int_0^1 du_1 \int_0^{u_1} du_2 \dfrac{1}{1+(1-u_1)(u_1-u_2) q_1} \nonumber \\
    && + \int_0^1 du_2 \int_0^{u_2} du_1 \dfrac{1}{1+(u_2-u_1) u_1 q_1} \Bigg). \label{T_w1_sectors}
\end{eqnarray}

Performing the change of variables $u_i \rightarrow 1-u_i$ in the first integral of the equation \eqref{T_w1_sectors}, the two integration sectors becomes the same. Hence, the equation \eqref{T_w1_sectors} simplifies to,  
\begin{eqnarray}
    T(q_1,0,0) &=& \dfrac{2}{m^2} \int_0^1 du_2 \int_0^{u_2} du_1 \dfrac{1}{1+(u_2-u_1) u_1 q_1}. \label{T_w1_oneintegral}
\end{eqnarray}
Interchanging the order of integration, and performing the integral over $u_2$, the expression for $T(q_1,0,0)$ above can be rewritten as, 
\begin{eqnarray}
    T(q_1,0,0) &=& \dfrac{2}{m^2 q_1} \left[G(0,r_+;1) + G(0,r_-;1) \right], \label{T_w1_G}    
\end{eqnarray}
where
\begin{eqnarray}
    r_{\pm}(q_1) = \dfrac{1}{2} \pm \dfrac{\sqrt{1+\frac{4}{q_1}}}{2}, \label{def_rpm_T}
\end{eqnarray}
and the functions $G(a,b;y)$ correspond to multiple polylogarithms of weight two. In general for weight $n$, the multiple polylogarithms are defined recursively as \cite{MultiplePolylogs},
\begin{eqnarray}
    G(a_1,\dots,a_n;y) = \int_0^y \dfrac{dx}{x-a_1}\, G(a_2,\dots,a_n;x), \label{def_Gs}
\end{eqnarray}
with 
\begin{eqnarray}
    G(a;y) &=& \ln \left( 1-\dfrac{y}{a} \right), \nonumber \\
    G(0;y) &=& \ln(y). \label{def_Gw1}
\end{eqnarray}

From the following identity for multiple polylogarithms of weight two \cite{MultiplePolylogs},
\begin{eqnarray}
    G(a,b;y) = \text{Li}_2\left(\dfrac{b-y}{b-a}\right) - \text{Li}_2\left(\dfrac{b}{b-a}\right) + \ln\left(1-\dfrac{y}{b}\right) \ln\left(\dfrac{y-a}{b-a}\right), \label{Iden_Gw2}
\end{eqnarray}
where the function $\text{Li}_2(x)$ represents a dilogarithm, the equation \eqref{T_w1_G} rewrite as,
\begin{eqnarray}
    T(w_1,0,0) &=& -\dfrac{2}{m^2 w_1} \Bigg\{ \dfrac{\pi^2}{3} + \ln\left[-\dfrac{r_-(q_1)}{r_+(q_1)}\right]\ln\left[\dfrac{r_-(q_1)}{r_+(q_1)}\right] - \text{Li}_2\left[-\dfrac{r_+(q_1)}{r_-(q_1)}\right] \nonumber \\
    && - \text{Li}_2\left[-\dfrac{r_-(q_1)}{r_+(q_1)}\right] \Bigg\}. \label{T_w1_dilogs_logs} 
\end{eqnarray} 
Setting $z=-r_+(q_1)/r_-(q_1)$, and using the well known dilogaritm identity,
\begin{eqnarray}
    \Li(z)+\Li\left(\frac{1}{z}\right) = - \dfrac{1}{2} \ln^2(-z) - \dfrac{\pi^2}{6},
\end{eqnarray}
the following final form for $T(q_1,0,0)$ is acquired,
\begin{eqnarray}
    T(q_1,0,0) = \dfrac{1}{m^2 q_1} \overline{T}(-m^2q_1^2), \label{T_finalform}
\end{eqnarray}
where
\begin{equation}
    \overline{T}(q) = \ln^2\left( \dfrac{\beta_{\hat{q}}-1}{\beta_{\hat{q}}+1} \right), \label{def_overT}
\end{equation}
with $\hat{q}=q/(4m^2)$, and the function $\beta_x$ being defined in the equation \eqref{def_beta_functions} of the appendix \ref{Disc_Box_Append}.

The result of the equation \eqref{T_finalform} agrees with the formula given in \cite{Britto_triangle_diagrams}, provided that all masses are assumed to be equal, and the following identifications are made: $w_1=r_+$, $\bar{w}_1=r_-$, along with the corresponding adjustments in the conventions for the spacetime metric.

The expression for $T(q_1,0,0)$ in the equation \eqref{T_finalform} presents a discontinuity across a branch cut with respect to the kinematical external invariant $q_1$. In order to compute this discontinuity, the following operator is defined \cite{DiscFeynmanInt},
\begin{eqnarray}
    \text{Disc}_s F(s) = \lim_{\epsilon \rightarrow 0} \left[ F(s+i\epsilon) - F(s-i\epsilon) \right]. \label{def_Disc}
\end{eqnarray}
From this definition, $\text{Disc}_s F(s)$ represents the discontinuity of $F$ as the variable $s$ crosses the real axis. 

In our case for the evaluation of $\dsc_{q_1} T(q_1,0,0)$, the kinematical region for the discontinuity is fixed such that $q_1 < -4$. Thus, from the definition given the equation \eqref{def_Disc}, the application of the discontinuity operator to the integral representation of $T(q_1,0,0)$ given in the equation \eqref{T_w1_oneintegral} reads as,
\begin{eqnarray}
    \dsc_{q_1} T(q_1,0,0) =- \dfrac{4i}{m^2} \lim_{\epsilon \rightarrow 0} \int_0^1 du_2 \int_0^{u_2} du_1 \dfrac{ \Lambda_{q_1} \epsilon }{ \bar{\Lambda}^2 + \Lambda_{q_1}^2 \epsilon^2  }, \label{DiscT_doubleintegral}
\end{eqnarray}
where 
\begin{eqnarray}
    \bar{\Lambda} = 1+(u_2-u_1) u_1 q_1, \qquad
    \Lambda_{q_1} = (u_2-u_1) u_1. \label{def_LambdasT}
\end{eqnarray}
Here, the integral representation for $T(q_1,0,0)$ is preferred over the final form presented in Equation \eqref{T_finalform}, as the subsequent section focuses solely on the discontinuities of the worldline integrals. From these discontinuities, the cut-constructible portion of the amplitudes is then inferred.

Now, from the well-know Sokhotski–Plemelj formula
\begin{eqnarray}
    \lim_{\epsilon\rightarrow0} \dfrac{1}{x \pm i \epsilon} = \text{PV}\dfrac{1}{x} \mp i \pi \delta(x), \label{Plemelj_formula}
\end{eqnarray}
it can be readily demonstrated that
\begin{eqnarray}
   \lim_{\epsilon \rightarrow 0} \dfrac{\epsilon}{x^2+\epsilon^2} = \pi\, \delta(x). \label{DiracDelta_alternative_def}
\end{eqnarray}
Using the expression above in the equation \eqref{DiscT_doubleintegral}, the discontinuity $\dsc_{q_1} T(q_1,0,0)$ rewrite as,
\begin{eqnarray}
    \dsc_{q_1} T(q_1,0,0) &=& - \dfrac{4i\pi}{m^2}  \int_0^1 du_2 \int_0^{u_2} du_1 \dfrac{1}{\Lambda_{q_1}} \delta\left( \dfrac{\bar{\Lambda}}{\Lambda_{q_1}} \right).
    \label{DiscT_delta}
\end{eqnarray}

The integral over $u_1$ in the equation \eqref{DiscT_delta} can be evaluated employing the following identity for the Dirac delta function,
\begin{eqnarray}
    \delta\left[ g(x) \right] = \sum_{i=1}^N \dfrac{\delta(x-x_i)}{|g^\prime(x_i)|}, \label{DiracDelta_composite}
\end{eqnarray} 
where $g(x_i) =0$. Thus, from the above identity, the expression for $\dsc_{q_1} T(q_1,0,0)$ in the equation \eqref{DiscT_delta} becomes,  
\begin{eqnarray}
    \dsc_{q_1} T(q_1,0,0) &=& - \dfrac{8i\pi}{m^2} \int_{\frac{2}{\sqrt{- q_1}}}^{1}  \dfrac{du_2}{\sqrt{u_2^2+\frac{4}{q_1}}} ,
    \label{DiscT_oneintegral}
\end{eqnarray}
where it has been assumed that $q_1 < -4$ for the evaluation of the Dirac delta function. Note that in this kinematical regime, only one root of $\bar{\lambda}=0$ contributes to the double integral of the equation \eqref{DiscT_delta}, and the lower limit of the integral over $u_2$ is modified by the constraints imposed by the delta function $\delta\left(\frac{\bar{\Lambda}}{\Lambda_{q_1}}\right)$.   

After evaluating the remaining integral in the equation \eqref{DiscT_oneintegral}, the expression for $\dsc_{q_1} T(q_1,0,0)$ reads as,
\begin{eqnarray}
    \dsc_{q_1} T(q_1,0,0) = \dfrac{4i\pi}{m^2 q_1} \ln\left( -z \right), \label{DiscT_Final}
\end{eqnarray}
with $z=-r_+(q_1)/r_-(q_2)$ as defined previously. This expression agrees with the result presented in \cite{Book_Weinzierl} for one-mass triangle integrals.

In the next section the discontinuities for the worldline integrals that arise in the evaluation of four-photon amplitudes within scalar/spinor QED are computed by following the same procedure outlined here, starting by direct application of the discontinuity operator given in the equation \eqref{def_Disc}. 

\section{Discontinuities of four-point one-loop worldline integrals onshell} \label{Four_point_wl_integrals}

In the case of the one-loop four-photon amplitude with on-shell photons in scalar and spinor QED, the master formulas of the worldline formalism yields the following integrals in $D=4$-dimensions \cite{WL_QED3,WL_fourphoton} (compare with equation \eqref{wl_integrals}),
\begin{eqnarray}
    \Gamma[\dG] = \int_0^1 \left( \prod_{i=1}^4 du_i \right) \dfrac{P(\dG)}{m^4\left(1+\Lambda^{(4)}\right)^2}, \label{wl_int_4photon}
\end{eqnarray}
where $P$ is a polynomial function on $\dG$ which is independent of the external kinematical invariants, and
\begin{eqnarray}
    \Lambda^{(4)} = \dfrac{1}{2} \left[ \bar{s}(G_{12}+G_{34}-G_{14}-G_{23}) + \bar{t}(G_{13}+G_{24}-G_{14}-G_{23}) \right]. \label{Lambda_4ph}
\end{eqnarray}
Here the Green's functions $G_{ij}$ and its derivatives are defined in the equation \eqref{G_def_scaled}.

The variable $\bar{s}$ and $\bar{t}$ in the equation \eqref{Lambda_4ph} represents normalized Mandelstam variables defined as,
\begin{eqnarray}
    \bar{s} = \frac{s}{m^2}, \qquad \bar{t} = \frac{t}{m^2}, \label{norm_Mandelstam}
\end{eqnarray}
with $s$, $t$ and $u$ given by
\begin{eqnarray}
    s &=& - (p_1+p_2)^2 = - 2 p_1 \cdot p_2 = - 2 p_3 \cdot p_4, \nonumber\\
    t &=& -(p_1+p_3)^2 = - 2 p_1 \cdot p_3 = -2 p_2 \cdot p_4, \nonumber \\
    u &=& -(p_1+p_4)^2 = - 2p_1 \cdot p_4 = -2 p_2 \cdot p_3, \label{Mandelstam}
\end{eqnarray}
which fulfill 
\begin{eqnarray}
    s + t + u = 0. \label{Mandelstam_constraint}  
\end{eqnarray}
The momenta $p_i$ in the equation \eqref{Mandelstam} corresponds to the external photons of the amplitude which are assumed to be onshell.

In the next section the discontinuity of the worldline integral $\Gamma[\dG]$ in the $s$ channel is computed by a direct application of the discontinuity operator given in the equation \eqref{def_Disc}. 

\subsection{Discontinuity in the $\bar{s}$ variable }

In order to apply the discontinuity operator of the equation \eqref{def_Disc} in the variable $\bar{s}$, assume the following kinematical region: $\bar{t} < 0$ and $\bar{s}<0$, with $|\bar{s}|>1>|\bar{t}|$ and $\bar{u}>4$.   

Now from equations \eqref{def_Disc} and \eqref{wl_int_4photon}, the discontinuity $\dsc_{\bar{s}} \Gamma[P(\dG)]$ reads as,
\begin{eqnarray}
    \dsc_{\bar{s}} \Gamma[P(\dG)] = -\dfrac{4i}{m^4} \lim_{\epsilon \rightarrow 0} \int_0^1 \left( \prod_{i=1}^4 du_i \right) \dfrac{ \epsilon P(\dG) F_{\bar{s}} }{\Lambda^{(4)}_{\bar{s}}(F_{\bar{s}}^2+\epsilon^2)^2} ,   \label{discGamma_4photon_4int}
\end{eqnarray}
where 
\begin{eqnarray}
    F_{\bar{s}} &=& \dfrac{1+\Lambda^{(4)}}{\Lambda^{(4)}_{\bar{s}}}, \nonumber \\
    \Lambda^{(4)}_{\bar{s}} &=& \dfrac{1}{2}(G_{12}+G_{34}-G_{14}-G_{23}). \label{def_F_Lambdas}
\end{eqnarray}

The discontinuity $\dsc_{\bar{s}} \Gamma[\dG]$ in the equation \eqref{discGamma_4photon_4int} can be evaluated following the same strategy outlined in the section \ref{One-mass_triangle_section}, where one of the worldline parameters $u_i$ is set to zero splitting the remaining integrals into all possible integration sectors. Following this approach, the discontinuity $\dsc_{\bar{s}} \Gamma[\dG]$ rewrite as,  
\begin{eqnarray}
    \dsc_{\bar{s}} \Gamma[P(\dG)] = \sum_{i=1}^6 \Gamma_{\bar{s}}^{(i)}(\bar{s},\bar{t}), \label{disc_4ph_s_Gammas}
\end{eqnarray}
where the functions $\Gamma_{\bar{s}}^{(i)}$ correspond to, 
\begin{eqnarray}
    \Gamma_{\bar{s}}^{(i)}(\bar{s},\bar{t}) &=& \dfrac{2i}{m^4} \lim_{\epsilon \rightarrow 0} \, \int_0^1 dx \int_0^x dy \int_0^y dz \left(\dfrac{P^{(i)}(x,y,z)}{\left[\Lambda_{\bar{s}}^{(4, i)}(x,y,z)\right]^2} \right) \nonumber \\
    && \times \dfrac{d}{dF^{(i)}}\left( \dfrac{\epsilon}{[F^{(i)}]^2+\epsilon^2}\right). \label{SectorIntegrals_s}
\end{eqnarray}
Here, the remaining integration variables of the equation \eqref{discGamma_4photon_4int}, after following the procedure described above, have been renamed to $x$, $y$ and $z$. The label $i$ in $\Gamma_{\bar{s}}^{(i)}$ stands for the corresponding integration sector, which after setting $u_4=0$ in equation \eqref{discGamma_4photon_4int}, are defined as, 
\begin{enumerate}
    \item $0\leq u_3 \leq u_2 \leq u_1 \leq 1 $.
    \item $0\leq u_3 \leq u_1 \leq u_2 \leq 1 $.
    \item $0\leq u_2 \leq u_3 \leq u_1 \leq 1 $.
    \item $0\leq u_2 \leq u_1 \leq u_3 \leq 1 $.
    \item $0\leq u_1 \leq u_2 \leq u_3 \leq 1 $.
    \item $0\leq u_1 \leq u_3 \leq u_2 \leq 1 $.
\end{enumerate}

The functions $P^{(i)}$, $F^{(i)}$ and $\Lambda_{\bar{s}}^{(4, i)}$ in equation \eqref{SectorIntegrals_s} are derived from $P(\dG)$, $F$ and $\Lambda_{\bar{s}}^{(4)}$ defined in equation \eqref{discGamma_4photon_4int} and are evaluated in the corresponding sector of integration. 

Further progress in the evaluation of the integrals $\Gamma_{\bar{s}}^{(i)}$ can be made by applying the identity \eqref{Derivative_delta_identity} to the integrals given in the equation \eqref{SectorIntegrals_s}. For example, for the first sector $0\leq u_3 \leq u_2 \leq u_1 \leq 1$, the integral in the equation \eqref{SectorIntegrals_s} becomes,
\begin{eqnarray}
    \Gamma_{\bar{s}}^{(1)}(\bar{s},\bar{t}) &=& \dfrac{2i}{m^4} \lim_{\epsilon \rightarrow 0} \, \int_0^1 dx \int_0^x dy \int_0^y dz \left(\dfrac{P^{(1)}(x,y,z)}{\left[z-y(1-x+z)\right]^2}\right) \nonumber \\
    && \times \dfrac{d}{dF^{(1)}}\left( \dfrac{\epsilon}{[F^{(1)}]^2+\epsilon^2}\right), \label{Gamma_s_1}
\end{eqnarray}
with 
\begin{eqnarray}
    F^{(1)}(x,y,z) &=& \dfrac{1-\bar{s} y (1-x)+z(\bar{s}+\bar{t}x+\bar{u}y)}{z-y(1-x+z)}, \nonumber \\
    P^{(1)}(x,y,z) &=& \left.P(\dG)\right\vert_{u_4=0, u_3=z,u_2=y,u_1=x; \, 0 \leq z \leq y \leq x \leq 1} . \label{F_s_1_P1}
\end{eqnarray}

To apply equation \eqref{Derivative_delta_identity}, the roots of $F^{(1)}(x,y,z)$ with respect to the variable $z$ are required. Therefore, from the condition $F^{(1)}(x,y,z_0) = 0$, the following expression for $z^{(1)}_0$ is obtained, 
\begin{eqnarray}
    z^{(1)}_0 = \dfrac{1-\bar{s}(1-x)y}{-\bar{s}(1-y)-\bar{t}(x-y)}. \label{root_z_Gamma_s_1}
\end{eqnarray}

If $\Gamma_{\bar{s}}^{(1)}$ in the equation \eqref{Gamma_s_1} does not vanish, the value of $z_0$ given in the equation \eqref{root_z_Gamma_s_1} must lie within the integration region $0\leq z_0 \leq y$ subject to the conditions $0\leq y \leq x$ and $0\leq x \leq 1$. From these inequalities, the following relations are obtained,
\begin{eqnarray}
    R_- \leq y \leq R_+\,, \qquad  \dfrac{2}{\sqrt{\bar{u}}} \leq x \leq 1, \label{Gamma_s_1_new_integrationlimits}
\end{eqnarray}
where
\begin{eqnarray}
    R_{\pm} = \dfrac{1}{2} \left( x \pm \sqrt{x^2-\dfrac{4}{\bar{u}}} \right). \label{def_Rpm}
\end{eqnarray}

Putting the pieces together, and after using the identity \eqref{Derivative_delta_identity}, the expression for $\Gamma_{\bar{s}}^{(1)}$ in the equation \eqref{Gamma_s_1} becomes,
\begin{eqnarray}
    \Gamma_{\bar{s}}^{(1)}(\bar{s},\bar{t}) &=& \dfrac{2i\pi}{m^4} \Bigg\{ \int_0^1 dx \int_0^x dy \Bigg[ \dfrac{\left.P^{(1)}(x,y,z)\right\vert_{z=y}}{ y + \bar{t} (1-x)(y-x) y - 1   } \delta\left( \dfrac{1}{(x-y)y}-\bar{u} \right) \nonumber \\
    &&  - \dfrac{\left.P^{(1)}(x,y,z)\right\vert_{z=0}}{ y + \bar{t} (1-x)(y-x) y - 1   } \delta\left( \bar{s} - \dfrac{1}{(1-x)y} \right) \Bigg] \nonumber \\
    && + \int_{ 2/\sqrt{\bar{u}} }^1 dx \int_{R_-}^{R_+} dy \left.\dfrac{dP^{(1)}(x,y,z)}{dz}\right\vert_{z=z_0^{(1)}}\left[\dfrac{1}{ (\bar{s}+\bar{t} x + \bar{u} y)^2 }\right] \Bigg\}. \label{Gamma_s_1_DiracDeltas}
\end{eqnarray}
Performing a change in the order of integration in the first integral of the equation \eqref{Gamma_s_1_DiracDeltas}, the expression for $\Gamma_{\bar{s}}^{(1)}$ can be rewritten as, 
\begin{eqnarray}
    \Gamma_{\bar{s}}^{(1)}(\bar{s},\bar{t}) &=& \dfrac{2i\pi}{m^4} \Bigg\{ \int_0^1 dy \int_y^1 dx  \dfrac{\left.P^{(1)}(x,y,z)\right\vert_{z=y}}{ y + \bar{t} (1-x)(y-x) y - 1   } \delta\left( \dfrac{1}{(x-y)y}-\bar{u} \right) \nonumber \\
    &&  - \int_0^1 dx \int_0^x dy \dfrac{\left.P^{(1)}(x,y,z)\right\vert_{z=0}}{ y + \bar{t} (1-x)(y-x) y - 1   } \delta\left( \bar{s} - \dfrac{1}{(1-x)y} \right)  \nonumber \\
    && + \int_{ 2/\sqrt{\bar{u}} }^1 dx \int_{R_-}^{R_+} dy \left.\dfrac{dP^{(1)}(x,y,z)}{dz}\right\vert_{z=z_0^{(1)}}\left[\dfrac{1}{ (\bar{s}+\bar{t} x + \bar{u} y)^2 }\right] \Bigg\}. \label{Gamma_s_1_DiracDeltas_after_change_order}
\end{eqnarray}

The evaluation of the Dirac delta functions in the equation \eqref{Gamma_s_1_DiracDeltas_after_change_order} does not only change the integrand, but also the integration limits of the remaining integral. For the case of the first double integral of the equation \eqref{Gamma_s_1_DiracDeltas_after_change_order}, the Dirac delta function states that,
\begin{eqnarray}
    \dfrac{1}{(x-y)y}-\bar{u} = 0,
\end{eqnarray}
which after solving for $x$, leads to,
\begin{eqnarray}
    x = y + \dfrac{1}{\bar{u} y}.
\end{eqnarray} 
In order to have a non-zero evaluation after applying the Dirac delta function, the expression above for $x$ must be compatible with the region of integration. Thus,
\begin{eqnarray}
    y \leq y + \dfrac{1}{\bar{u} y} \leq 1, \qquad 0 \leq y \leq 1.
\end{eqnarray}
From these inequalities, the variable $y$ is restricted to lie within the following region of integration,
\begin{eqnarray}
    r_{-}(-\bar{u}) \leq y \leq r_+(-\bar{u}),
\end{eqnarray}
where $r_{\pm}(a)$ is defined in the equation \eqref{def_rpm_T}.

Analogously, the second double integral of the equation \eqref{Gamma_s_1_DiracDeltas_after_change_order} states that,
\begin{eqnarray}
    \bar{s} - \dfrac{1}{(1-x)y} = 0,
\end{eqnarray}
which after solving for $y$ produce the equality,
\begin{eqnarray}
    y = \dfrac{1}{\bar{s}(1-x)}.
\end{eqnarray}
However, this result is incompatible with the region of integration for $\bar{s} \leq 0$ (which corresponds to the kinematical region where the discontinuity is evaluated). Hence, the second double integral in the equation \eqref{Gamma_s_1_DiracDeltas_after_change_order} vanishes. 

Thus, after evaluation of the Dirac delta functions in the equation \eqref{Gamma_s_1_DiracDeltas_after_change_order}, the expression for $\Gamma_{\bar{s}}^{(1)}$ can be rewritten as,
\begin{eqnarray}
    \Gamma_{\bar{s}}^{(1)}(\bar{s},\bar{t}) &=& \dfrac{2i\pi}{m^4} \Bigg\{ \int_{r_-(-\bar{u})}^{r_+(-\bar{u})} dy \dfrac{\left.P^{(1)}(x,y,z)\right\vert_{z=y,x=x^{(1)}_0}}{ \bar{t} + \bar{s} \bar{u} (1-y) y   } \nonumber \\
    && + \int_{ 2/\sqrt{\bar{u}} }^1 dx \int_{R_-}^{R_+} dy \left.\dfrac{dP^{(1)}(x,y,z)}{dz}\right\vert_{z=z_0^{(1)}}\left[\dfrac{1}{ (\bar{s}+\bar{t} x + \bar{u} y)^2 }\right] \Bigg\}, \label{Gamma_s_1_EliminationDiracDeltas}
\end{eqnarray}
where 
\begin{eqnarray}
    x^{(1)}_0 = y + \dfrac{1}{\bar{u} y}. \label{def_x01}
\end{eqnarray}

Finally, interchanging the order of integration in the second double integral of the equation \eqref{Gamma_s_1_EliminationDiracDeltas}, the firs term $\Gamma_{\bar{s}}^{(1)}$ of the discontinuity $\dsc_{\bar{s}} \Gamma[\dG]$ simplifies to,
\begin{eqnarray}
    \Gamma_{\bar{s}}^{(1)}(\bar{s},\bar{t}) &=& \dfrac{2i\pi}{m^4} \Bigg\{ \int_{r_-(-\bar{u})}^{r_+(-\bar{u})} dy \Bigg[ \dfrac{\left.P^{(1)}(x,y,z)\right\vert_{z=y,x=x^{(1)}_0}}{ \bar{t} + \bar{s} \bar{u} (1-y) y   } \nonumber \\
    && + \int_{ x_0^{(1)} }^1 dx\, \left.\dfrac{dP^{(1)}(x,y,z)}{dz}\right\vert_{z=z_0^{(1)}}\left[\dfrac{1}{ (\bar{s}+\bar{t} x + \bar{u} y)^2 }\right] \Bigg\}. \label{Gamma_s_1_Final}
\end{eqnarray}

Following the same procedure for the other sectors, the expressions for the remaining $\Gamma_{\bar{s}}^{(i)}$ read as,
\begin{eqnarray}
    \Gamma_{\bar{s}}^{(2)}(\bar{s},\bar{t}) &=& 0, \nonumber \\
    \Gamma_{\bar{s}}^{(3)}(\bar{s},\bar{t}) &=& \dfrac{2i\pi}{m^4} \Bigg\{ \int_{r_-(-\bar{u})}^{r_+(-\bar{u})} dy \Bigg[ \dfrac{\left.P^{(3)}(x,y,z)\right\vert_{z=y,x=x^{(1)}_0}}{ \bar{s} + \bar{t} \bar{u} (1-y) y   } \nonumber \\
    && + \int_{ x_0^{(1)} }^1 dx\, \left.\dfrac{dP^{(3)}(x,y,z)}{dz}\right\vert_{z=z_0^{(3)}}\left[\dfrac{1}{ (\bar{t}+\bar{s} x + \bar{u} y)^2 }\right] \Bigg\} , \nonumber \\
    \Gamma_{\bar{s}}^{(4)}(\bar{s},\bar{t}) &=& 0, \nonumber \\
    \Gamma_{\bar{s}}^{(5)}(\bar{s},\bar{t}) &=& -\dfrac{2i\pi}{m^4} \Bigg\{ \int_{r_-(-\bar{u})}^{r_+(-\bar{u})} dx \Bigg[ \dfrac{\left.P^{(5)}(x,y,z)\right\vert_{z=0,y=y^{(5)}_0}}{ \bar{s} + \bar{u}^2(1-x)^2-\bar{s}\bar{u}(1-x)x   } \nonumber \\
    && + \int_{ y_0^{(5)} }^x dy\, \left.\dfrac{dP^{(5)}(x,y,z)}{dz}\right\vert_{z=z_0^{(5)}}\left[\dfrac{1}{ [\bar{t}(1-x)+\bar{s}(1-y)]^2 }\right] \Bigg\}, \nonumber \\
    \Gamma_{\bar{s}}^{(6)}(\bar{s},\bar{t}) &=& -\dfrac{2i\pi}{m^4} \Bigg\{ \int_{r_-(-\bar{u})}^{r_+(-\bar{u})} dx \Bigg[ \dfrac{\left.P^{(6)}(x,y,z)\right\vert_{z=0,y=y^{(5)}_0}}{ \bar{t} + \bar{u}^2(1-x)^2-\bar{t}\bar{u}(1-x)x   } \nonumber \\
    && + \int_{ y_0^{(5)} }^x dy\, \left.\dfrac{dP^{(6)}(x,y,z)}{dz}\right\vert_{z=z_0^{(6)}}\left[\dfrac{1}{ [\bar{s}(1-x)+\bar{t}(1-y)]^2 }\right] \Bigg\}, \label{Gamma_s_i}
\end{eqnarray}
where $x_0^{(1)}$ is defined in the equation \eqref{def_x01} and,
\begin{eqnarray}
    y^{(5)}_0 &=& \dfrac{1}{\bar{u}(1-x)}, \\
    z^{(3)}_0 &=& \dfrac{1-\bar{t}(1-x)y}{-\bar{t}(1-y)-\bar{s}(x-y)}, \nonumber \\
    z^{(5)}_0 &=& \dfrac{\bar{u}(1-x)y-1}{-\bar{t}(1-x)-\bar{s}(1-y)}, \nonumber \\
    z^{(6)}_0 &=& \dfrac{\bar{u}(1-x)y-1}{-\bar{s}(1-x)-\bar{t}(1-y)}, \nonumber \\
    P^{(3)} &=& \left.P(\dG)\right\vert_{u_4=0, u_2=z,u_3=y,u_1=x; \, 0 \leq z \leq y \leq x \leq 1}, \nonumber \\
    P^{(5)} &=& \left.P(\dG)\right\vert_{u_4=0, u_1=z,u_2=y,u_3=x; \, 0 \leq z \leq y \leq x \leq 1}, \nonumber \\
    P^{(6)} &=& \left.P(\dG)\right\vert_{u_4=0, u_1=z,u_3=y,u_2=x; \, 0 \leq z \leq y \leq x \leq 1}.    
    \label{aux_def_Gammas}
\end{eqnarray}

\section{Examples of the application of the discontinuity method} \label{Disc_method_examples}

In this section, the master formulas derived previously for the computation of discontinuities are used as a tool to evaluate a couple of examples of four-point one-loop worldline integrals onshell. The first example that will be considered is the simplest case where $P(\dG) = 1$ in equation \eqref{wl_int_4photon}, which corresponds to the one-loop scattering amplitude in the $\phi^3$ field theory for four external scalar particles. 

In the following, for the evaluation of the discontinuities, it will always be assumed the kinematical region: $\bar{s}<-1<\bar{t}<0$ with $\bar{u}>4$.

\subsection{Evaluation of one-loop scattering amplitude in the $\phi^3$ field theory for four external scalar particles}

The four-point one-loop amplitude for the $\phi^3$ field theory is proportional to the worldline master formula given in the equation \eqref{wl_int_4photon} by setting $P(\dG) = 1$,
\begin{eqnarray}
    \Gamma(s,t,u) = \int_0^1 \left( \prod_{i=1}^4 du_i \right) \dfrac{1}{m^4[1+\Lambda^{(4)}(s,t,u)]^2}, \label{amplitude_4_scalar}
\end{eqnarray}
where to make explicit the dependence in the Mandelstam variables, the function $\Lambda^{(4)}(s,t,u)$ can be rewritten as,
\begin{eqnarray}
    \Lambda^{(4)}(s,t,u) = \dfrac{1}{2}\left[ \bar{s}(G_{12}+G_{34})+ \bar{t}(G_{13}+G_{24}) + \bar{u}(G_{14}+G_{23}) \right]. 
\end{eqnarray}

Note that the function $\Gamma(s,t,u)$ in the equation \eqref{amplitude_4_scalar} is invariant under any permutation of the Madelstam variables.

Now, according to the equations \eqref{disc_4ph_s_Gammas}, \eqref{Gamma_s_1} and \eqref{Gamma_s_i}, the discontinuity $\dsc_{\bar{s}}\Gamma(s,t,u)$ reads as,
\begin{eqnarray}
    \dsc_{\bar{s}}\Gamma(s,t,u) &=& \dfrac{2i\pi}{m^4} \Bigg\{ \int_{r_-(-\bar{u})}^{r_+(-\bar{u})} dx \Bigg[ \dfrac{1}{ \bar{t} + \bar{s} \bar{u} (1-x) x   } + \dfrac{1}{ \bar{s} + \bar{t} \bar{u} (1-x) x   } \Bigg] \nonumber \\
    && -   \dfrac{1}{ \bar{s} + \bar{u}^2(1-x)^2-\bar{s}\bar{u}(1-x)x   } + \dfrac{1}{ \bar{t} + \bar{u}^2(1-x)^2-\bar{t}\bar{u}(1-x)x   } \Bigg] \Bigg\}. \nonumber \\ 
\end{eqnarray}
After integration, the equation above becomes,
\begin{eqnarray}
    \dsc_{\bar{s}}\Gamma(s,t,u) &=& \dfrac{2i\pi}{m^4} \Bigg\{ \dfrac{1}{\bar{s}\bar{u} \beta_{\hs\hu}} \Bigg[ 2\ln\left(\dfrac{\beta_{\hs\hu}+\beta_{\hu}}{\beta_{\hs\hu}-\beta_{\hu}}\right) \nonumber \\
    && + \ln\left( \dfrac{(u+t\beta_{\hu}-s\beta_{\hs\hu})(u-t\beta_{\hu}+s\beta_{\hs\hu})}{(u-t\beta_{\hu}-s\beta_{\hs\hu})(u+t\beta_{\hu}+s\beta_{\hs\hu})} \right)  \Bigg] + (s \leftrightarrow t) \Bigg\},
    \label{Disc_G_4s_v1}
\end{eqnarray}
where $(s \leftrightarrow t)$ indicates the presence of a second term similar to the first, with $s$ and $t$ interchanged. The functions $\beta_{x}$ and $\beta_{x y}$ are defined in equation \eqref{def_beta_functions} in the Appendix \ref{Disc_Box_Append}.   

using the definition of the $\beta$'s functions, the argument of the second logarithm of the square bracket in equation \eqref{Disc_G_4s_v1} simplifies to,
\begin{eqnarray}
  \dfrac{(u+t\beta_{\hu}-s\beta_{\hs\hu})(u-t\beta_{\hu}+s\beta_{\hs\hu})}{(u-t\beta_{\hu}-s\beta_{\hs\hu})(u+t\beta_{\hu}+s\beta_{\hs\hu})} &=& \dfrac{2-\bar{s} \beta_{\hu}(\beta_{\hs\hu}+\beta_{\hu}) }{2+\bar{s} \beta_{\hu}(\beta_{\hs\hu}-\beta_{\hu})}  \nonumber \\
  &=& \left(\dfrac{\beta_{\hs\hu}+\beta_{\hu}}{\beta_{\hs\hu}-\beta_{\hu}}\right)^2. \label{beta_iden_Log}
\end{eqnarray}

From the identity \eqref{beta_iden_Log}, the expression in equation \eqref{Disc_G_4s_v1} for the discontinuity $\dsc_{\bar{s}}\Gamma(s,t,u)$ yields
\begin{eqnarray}
    \dsc_{\bar{s}}\Gamma(s,t,u) &=& -\dfrac{8i\pi}{u} \Bigg[ \dfrac{1}{s \beta_{\hs\hu}} \ln\left(\dfrac{\beta_{\hs\hu}-\beta_{\hu}}{\beta_{\hs\hu}+\beta_{\hu}}\right)  + (s \leftrightarrow t) \Bigg].
    \label{Disc_G_4s}
\end{eqnarray}

Comparing the equation \eqref{Disc_G_4s} with the discontinuities of the functions $D(s,u)$ and $D(t,u)$ given in the equations \eqref{Disc_s_Dsu} and \eqref{Disc_s_Dtu}, and which are defined in the equation \eqref{Dav_fun}, the amplitude $\Gamma(s,t,u)$ must acquire the following form,
\begin{eqnarray}
    \Gamma(s,t,u) = 4 \left[ D(s,u) + D(t,u) + D(s,t) \right] + R(s,t,u),
\end{eqnarray}
where it has been included the function $D(s,t)$ by the permutation invariance of $\Gamma(s,t,u)$, and the function $R(s,t,u)$ corresponds to a rational function on the Mandelstam variables which does not present a discontinuity with respect to the variable $\bar{s}$. In this case, if the Mandelstam variables and the mass of the scalar particle are fixed to some kinematical value, by a numerical evaluation it can be shown that the rational function $R(s,t,u)$ vanishes. Hence, the amplitude $\Gamma(s,t,u)$ corresponds to,
\begin{eqnarray}
    \Gamma(s,t,u) = 4\left[ D(s,u) + D(t,u) + D(s,t)\right]. \label{Gamma_4_fullscalar}
\end{eqnarray}

This example demonstrates how the worldline integrals can be evaluated, up to a rational term, by calculating the discontinuity of the integral. In the following section, a different example is presented, where the polynomial $P(\dG)$ is not constant. 

\subsection{Evaluation of a worldline integral for the one-loop four-photon amplitude of scalar QED  }

One of the principal integrals that arise in the evaluation of one-loop four-photon amplitudes of scalar QED in $D=4$ dimensions corresponds to \cite{WL_QED3,WL_fourphoton,WL_4photon_amplitudes}\footnote{A comprehensive analysis of the one-loop four-photon amplitude in both scalar and spinor QED within the worldline formalism can be found in Ref. \cite{WL_4photon_amplitudes}}, 
\begin{eqnarray}
    \Gamma_{(1234)}(s,t,u) = \int_0^1 \left( \prod_{i=1}^4 du_i \right) \dfrac{\dG_{12}\dG_{23}\dG_{34}\dG_{41}}{m^4[1+\Lambda^{(4)}(s,t,u)]^2}, \label{cyclic_int_4u}
\end{eqnarray}
which contains a cyclic polynomial $P(\dG)$ in the integration variables $u_{1,2,3,4}$. Following a similar strategy used in the unitarity methods of the standard formalism of quantum field theory, where a one-loop amplitude can be rewritten as a linear combination of tadpoles, bubble, triangle and box integrals together with a rational function, here the following ansatz is proposed for $\Gamma_{(1234)}(s,t,u)$,
\begin{eqnarray}
    \Gamma_{(1234)}(s,t,u) &=&  r_0 (s,t,u) + r_1(s,t,u) \overline{B}(s) + r_2(s,t,u) \overline{B}(t) + r_3(s,t,u) \overline{B}(u) \nonumber \\
    && + r_4(s,t,u) \overline{T}(s) + r_5(s,t,u) \overline{T}(t) + r_6(s,t,u) \overline{T}(u) \nonumber \\
    && + r_7(s,t,u) D(s,t) + r_8(s,t,u) D(s,u) + r_9(s,t,u) D(t,u), \label{Gcycl_ansatz}
\end{eqnarray}
where $r_i(s,t,u)$ are rational functions in the Mandelstam variables, and $\overline{B}(x)$, $\overline{T}(x)$, $D(x,y)$ are the transcendental functions associated with the bubble, triangle and box integrals defined in the equations \eqref{Bubblebar}, \eqref{def_overT} and \eqref{Dav_fun}, respectively. Since $\Gamma_{(1234)}(s,t,u)$ is free of ultraviolet divergences, the ansatz in equation \eqref{Gcycl_ansatz} contains only UV-finite contributions of the bubble integrals in $D=4-2\epsilon$ dimensions, and any contributions from the tadpole integrals.  

In Ref. \cite{WL_fourphoton} is shown that the worldline integrals for the one-loop four-photon amplitude has the same structure given in the ansatz \eqref{Gcycl_ansatz}, provided that the function $\bar{B}(x,y)$ defined there will be proportional to the function $D(x,y)$ given in the equation \eqref{Dav_fun}. In fact, by numerical evaluation for example, it can be shown that,
\begin{eqnarray}
    \bar{B}(x,y) = 4 D(x,y). \label{Bbar_Dav_fun}
\end{eqnarray} 

Now, the rational functions $r_i(s,t,u)$ in the equation \eqref{Gcycl_ansatz}, with exception of the case when $i=0$, can be obtained from the discontinuity method, since from equation \eqref{Bubblebar},
\begin{eqnarray}
    \dsc_{\bar{s}} \overline{B}(s) = 0, \quad \dsc_{\bar{s}} \overline{B}(t) = 0, \quad \dsc_{\bar{s}} \overline{B}(u) = - 2\pi i \beta_{\hu}, \label{disc_Bubblebar} 
\end{eqnarray}
and from equations \eqref{T_finalform} and \eqref{DiscT_Final},
\begin{eqnarray}
    \dsc_{\bar{s}} \overline{T}(s)=0, \quad \dsc_{\bar{s}} \overline{T}(t)=0, \quad \dsc_{\bar{s}} \overline{T}(u) = 4\pi i \ln\left( \dfrac{1+\beta_{\hu}}{1-\beta_{\hu}} \right). \label{disc_Trianglebar}
\end{eqnarray}
Additionally, the discontinuity with respect to $\bar{s}$ for the functions $D(x,y)$ are given in the equations \eqref{Disc_s_Dst}, \eqref{Disc_s_Dsu} and \eqref{Disc_s_Dtu} of the Appendix \ref{Disc_Box_Append}.

Thus, from the formulas given in the equations \eqref{Gamma_s_1_Final} and \eqref{Gamma_s_i}, with $P(\dG) = \dG_{12}\dG_{23}\dG_{34}\dG_{41} $, the discontinuity $\dsc_{\bar{s}}\Gamma_{(1234)}(s,t,u)$ reads as,
\begin{eqnarray}
    \dsc_{\bar{s}}\Gamma_{(1234)}(s,t,u) &=& - 2 \pi i \beta_{\hu} r_3(s,t,u) + 4\pi i r_6(s,t,u) \ln\left( \dfrac{1+\beta_{\hu}}{1-\beta_{\hu}} \right) \nonumber \\
     && - \dfrac{2\pi i}{su\beta_{\hs\hu}} r_8(s,t,u) \ln\left(\dfrac{\beta_{\hs\hu}-\beta_{\hu}}{\beta_{\hs\hu}+\beta_{\hu}}\right) \nonumber \\
     && - \dfrac{2\pi i}{tu\beta_{\hT\hu}} r_9(s,t,u) \ln\left(\dfrac{\beta_{\hT\hu}-\beta_{\hu}}{\beta_{\hT\hu}+\beta_{\hu}}\right)  ,
\end{eqnarray}
where 
\begin{eqnarray}
    r_3(s,t,u) &=& -32 \left( \dfrac{1}{3su} + \dfrac{s^2}{t^3u} + \dfrac{3s}{2t^2u}-\dfrac{1}{3tu}-\dfrac{t}{3s^2u} \right), \nonumber \\
    r_6(s,t,u) &=& 4 \left[ 4m^2 \left(\frac{1}{s^2 t}-\frac{2}{s u^2}-\frac{2}{t^3}\right)-\frac{4 t}{3 s^3}+\frac{4 s^2}{t^4}+\frac{1}{s^2}+\frac{4 s}{t^3}-\frac{4}{3 t^2} \right], \nonumber \\
    r_8(s,t,u) &=& -4 \Bigg[ 16 m^4 \left(\frac{s}{t^2 u}-\frac{2}{3 s u}+\frac{1}{t u}\right)+4m^2 \left(-\frac{8 s^3}{t^3 u}-\frac{16 s^2}{t^2 u}-\frac{14 s}{3 t u}+\frac{t}{s u}+\frac{10}{3 u}\right)\nonumber \\
    && -\frac{8 s^4}{t^4}-\frac{16 s^3}{t^3}-\frac{16 s^2}{3 t^2}+\frac{8 s}{3 t}+1 \Bigg], \nonumber \\
    r_9(s,t,u) &=& -4 \Bigg[ 16m^4 \left(\frac{s}{t^2 u}-\frac{2}{3 s u}+\frac{1}{t u}\right)+4m^2 \left(\frac{10 t^2}{3 s^2 u}+\frac{ t}{3 s u}-\frac{2 s}{t u}-\frac{4}{u}\right) \nonumber\\
    && -\frac{8 t^4}{3 s^3 u}-\frac{10 t^3}{3 s^2 u}+\frac{4 t^2}{3 s u}+\frac{s}{u}+\frac{3 t}{u} \Bigg].
\end{eqnarray}

Note that from the definition of $\Gamma_{(1234)}(s,t,u)$ in the equation \eqref{cyclic_int_4u},
\begin{eqnarray}
    \Gamma_{(1234)}(s,t,u) = \Gamma_{(1234)}(u,t,s).
\end{eqnarray}
which implies,
\begin{eqnarray}
    r_1(s,t,u) = r_3(u,t,s), \quad r_4(s,t,u) = r_6(u,t,s), \quad r_7(s,t,u) = r_9(u,t,s). 
\end{eqnarray}

To obtain the remaining rational functions $r_2$ and $r_5$ in equation \eqref{Gcycl_ansatz} that can be obtained from the discontinuity method, interchange $t \leftrightarrow u$ in equation \eqref{cyclic_int_4u} and perform the change of variables $u_3 \leftrightarrow u_4$. From this procedure, the polynomial $P(\dG)$ that will be used in the equations \eqref{Gamma_s_1_Final} and \eqref{Gamma_s_i} corresponds to,
\begin{eqnarray}
    P(\dG) = \dG_{12} \dG_{24} \dG_{43} \dG_{31}.
\end{eqnarray}
Using this polynomial in the discontinuity method, and interchange again $t \leftrightarrow u$ to reverse the previous changes, the following results for $r_2$ and $r_5$ are obtained,
\begin{eqnarray}
    r_2(s,t,u) &=& \dfrac{16}{3} \left( \frac{2}{s^2}+\frac{1}{s u}+\frac{2}{u^2} \right),  \nonumber \\
    r_5(s,t,u) &=& 4\left[ 4m^2 \left(-\frac{t^2}{s^2 u^3}-\frac{t}{s u^3}\right)+\frac{4 t^4}{3 s^3 u^3}+\frac{3 t^3}{s^2 u^3}+\frac{t^2}{s u^3}-\frac{2 s}{u^3}-\frac{4 t}{u^3} \right]. \nonumber \\
\end{eqnarray}

\section{Summary and conclusions} \label{Conclusion_summary}

The discontinuity method described in this work can be used as an efficient tool for the evaluation of integrals that emerge within the worldline formalism. As in the case of the unitarity methods of the standard diagrammatic approach of quantum field theory, the discontinuity of transcendental functions, as logarithms and dilogarithms, is exploited in order to circumvent the evaluation of otherwise difficult integrals. 

The discontinuity method for evaluating worldline integrals in one-loop amplitudes in $D = 4$ dimensions, as presented in Equation \eqref{wl_int_4photon}, involves rewriting the integrals as a linear combination of the transcendental functions $\overline{B}(x)$, $\overline{T}(x)$, and $D(x, y)$, which are defined in Equations \eqref{Bubblebar}, \eqref{def_overT} and \eqref{Dav_fun}, respectively. These functions arise in the evaluation of bubble, triangle, and box Feynman integrals, which are discussed throughout the article. Since each of these functions—$\overline{B}(x)$, $\overline{T}(x)$, and $D(x, y)$—has a characteristic discontinuity with respect to a given kinematical invariant, the rational functions multiplying them in the linear combination can be determined once the discontinuity of the worldline integral is computed. This task can be performed using the master formulas provided in Equations \eqref{Gamma_s_1_Final} and \eqref{Gamma_s_i}. Given that these formulas produce only logarithmic terms, they can be readily evaluated. 

As in the case of the unitarity methods in the standard formalism, rational functions that do not multiply any of the transcendental functions cannot be directly obtained from the discontinuity method. For these functions, an alternative method of evaluation is required. 

The application of the discontinuity method is illustrated in sections \ref{One-mass_triangle_section} and \ref{Disc_method_examples}. In Sec. \ref{One-mass_triangle_section} the framework for evaluating discontinuities in the worldline master integrals is presented, using the one-mass triangle integrals from Equation \eqref{Triangle_integral_one_mass} as an example. The role of the Dirac delta function, defined by the Sokhotski–Plemelj formula (Equation \eqref{Plemelj_formula}) and the change of the integration region in the evaluation of the discontinuities is also highlighted. In Sec. \ref{Disc_method_examples}, the discontinuity method is fully implemented using the master formulas fro Equations \eqref{Gamma_s_1_Final} and \eqref{Gamma_s_i} to evaluate, up to a rational function, worldline integrals that arise in the calculation of scattering amplitudes for four external particles in $\phi^3$ field theory and scalar QED. These examples demonstrate the potential of the discontinuity method as a powerful tool for evaluating worldline master integrals without the need for their direct computation. 

Although the QED examples presented in this article focus on the specific case of four-photon amplitudes, the discontinuity method is sufficiently general to be extended to an arbitrary number of photons.

\section*{Acknowledgments}
The author gratefully acknowledges C. Schubert for useful conversations and correspondence throughout the preparation of this manuscript, and {\em Secretaría de Ciencia, Humanidades, Tecnología e Innovación} SECIHTI (Mexico) for support.

\appendix 

\section{Conventions} \label{Conventions}

On the side of the worldline formalism, the Euclidean metric $(+,+,+,+)$ is used. On the field theory side, a Wick rotation is performed to Minkowski space with metric $\eta_{\mu\nu} = \text{diag}(-,+,+,+) $.  

\section{Gaussian integration for path integrals} \label{Integration_path_integrals}

Define the D-dimensional kinetic operator $K^{\mu\nu}$ and the vector $J^\mu$ as,
\begin{eqnarray}
    K^{\mu\nu} = -\delta^{\mu\nu} \dfrac{1}{2} \dfrac{d^2}{d\tau^2}, \qquad
    J^\mu = i \sum_{i=1}^N p_i^\mu \delta(\tau-\tau_i).  \label{Def_K_J}
\end{eqnarray}
where $\delta^{\mu\nu}$ is the Euclidean metric. Thus, the integral in the equation \eqref{scalar_Npoint_v2} formally evaluates to,
\begin{eqnarray}
    \int_{\text{DBC}} Dq\, \int_0^T \prod_{k=1}^N  d\tau_k  \, \e^{ \sum_{i=1}^N i p_i \cdot q(\tau_i) - \frac{1}{4} \int_0^T d\tau\, \dot{q}^2(\tau)  } && \nonumber \\
    && \hspace{-150pt} = N \, \left(\text{Det}\, K \right)^{-\frac{1}{2}} \e^{ -  \int_0^T d\tau d\tau^\prime\, \Delta(\tau,\tau^\prime) J(\tau) \cdot J(\tau^\prime) }, \label{path_integral_scalar}
\end{eqnarray}
where $N$ is a normalization factor and $-2\delta^{\mu\nu} \Delta(\tau,\tau^\prime)$ corresponds to the Green's function for the kinetic operator $K^{\mu\nu}$ with Dirichlet boundary conditions. 

The function $\Delta(\tau,\tau^\prime)$ reads as \cite{WL_notes},
\begin{eqnarray}
    \Delta(\tau,\tau^\prime) = \dfrac{\tau\tau^\prime}{T} + \dfrac{|\tau-\tau^\prime|}{2} - \dfrac{\tau+\tau^\prime}{2}, \label{Def_Delta_DBC}
\end{eqnarray}   
which satisfies the differential equation,
\begin{eqnarray}
    \dfrac{d^2}{d\tau^2} \Delta(\tau,\tau^\prime) = \delta(\tau-\tau^\prime),
\end{eqnarray}
with boundary conditions,
\begin{eqnarray}
    \Delta(0,\tau^\prime) = \Delta(T,\tau^\prime) = 0.
\end{eqnarray}

From the equations \eqref{Def_K_J} and \eqref{Def_Delta_DBC}, the exponential term in the right hand side of the equation \eqref{path_integral_scalar} rewrites as,
\begin{eqnarray}
    -  \int_0^T d\tau d\tau^\prime\, \Delta(\tau,\tau^\prime) J(\tau) \cdot J(\tau^\prime) && \nonumber \\
    && \hspace{-100pt} = \dfrac{1}{2} \sum_{i,j=1}^N p_i \cdot p_j \left[ G(\tau_i,\tau_j) + \dfrac{1}{T}(\tau_i^2+\tau_j^2) - (\tau_i+\tau_j) \right], \label{Iden_Js}
\end{eqnarray}
where the function $G(\tau,\tau^\prime)$ is defined in the equation \eqref{def_Gs_wl}. This function is proportional to the Green's function of the one-dimensional kinetic operator $K$ with the non-local boundary condition given by the equation \eqref{non_local_bc_q} (see for example \cite{Schubert_WL,WL_notes}). 

Now, in order to evaluate the determinant of the one-dimensional kinetic operator $K$, note that its eigenvalues $\lambda_n$ and eigenfunctions $q_n(\tau)$ with Dirichlet boundary conditions correspond to,
\begin{eqnarray}
    q_n(\tau) = \sqrt{\dfrac{2}{T}}\, \sin\left( \dfrac{n \pi}{T} \tau\right), \qquad \lambda_n = \dfrac{n^2 \pi^2}{2 T^2}, \qquad \text{for } n=1,2,3,\dots \label{eigenvalues_eigenfunctions_K}
\end{eqnarray}
The factor of $\sqrt{\frac{2}{T}}$ in the eigenfunctions is chosen such that they form an orthonormal basis in the interval $[0,T]$.

From the eigenvalues given in the equation \eqref{eigenvalues_eigenfunctions_K}, the determinant of the one-dimensional kinetic operator $K$ reads as,
\begin{eqnarray}
    \text{Det} \, K = \prod_{n=1}^\infty \left( \dfrac{n^2 \pi^2}{2 T^2} \right). \label{Det_K_product}
\end{eqnarray}
This ill-defined infinite product can be regularized using the $\zeta$-function regularization procedure, \cite{WL_notes, ZetaReg, Hawking_ZetaReg}. Applying this approach, the following identity is derived \cite{WL_notes}, 
\begin{eqnarray}
    \prod_{i=1}^\infty a n^b = \sqrt{\dfrac{(2\pi)^b}{a}}. \label{zeta_fun_reg_iden}
\end{eqnarray}

From the identity \eqref{zeta_fun_reg_iden}, the determinant of the one-dimensional kinetic operator $K$ in the equation \eqref{Det_K_product} becomes,
\begin{eqnarray}
    \text{Det} \, K = 2 \sqrt{2} T. \label{Det_K_product2} 
\end{eqnarray}
Thus, extending the above result to $D$-dimensions, together with the equation \eqref{Iden_Js}, the path integral in equation \eqref{path_integral_scalar} rewrite as,
\begin{eqnarray} 
    \int_{\text{DBC}} Dq\, \int_0^T \prod_{k=1}^N d\tau_k  \, \e^{ \sum_{i=1}^N i p_i \cdot q(\tau_i) - \frac{1}{4} \int_0^T d\tau\, \dot{q}^2(\tau)  } && \nonumber \\
    && \hspace{-150pt} = \left(4\pi\, T\right)^{-\frac{D}{2}} \e^{\frac{1}{2} \sum_{i,j=1}^N p_i \cdot p_j \left[ G(\tau_i,\tau_j) + \frac{1}{T}(\tau_i^2+\tau_j^2) - (\tau_i+\tau_j) \right]} , \label{path_integral_scalar2}
\end{eqnarray}
where the normalization factor $N$, which is independent of $T$, has been fixed such that the last result coincides with the non-relativistic free particle path integral.

\section{ Integral identity for the discontinuity of one-loop four-point worldline integrals on-shell} \label{DiracDelta_iden_Box}

In section \ref{Four_point_wl_integrals}, the following integrals emerge in the evaluation of the discontinuities of the four-point one-loop worldline master formulas, 
\begin{eqnarray}
  I = \lim_{\epsilon \rightarrow 0} \int_a^b du\, \phi(u) \, \dfrac{d}{dF}\left( \dfrac{\epsilon}{F^2+\epsilon^2} \right), \label{delta_int_disc_box} 
\end{eqnarray}
where $F=F(u)$, and it is assumed that in the interval $[a,b]$ there is only one root $u_0$ of the function $F(u)$ such that $F(u_0)=0$. 

After the change of variables $y=F(u)$ and integrating by parts, the integral $I$ in the equation \eqref{delta_int_disc_box} rewrite as,
\begin{eqnarray}
  I = \pi \left\{ \delta[F(b)] \dfrac{\phi(b)}{F^\prime(b)} - \delta[F(a)] \dfrac{\phi(a)}{F^\prime(a)} - \dfrac{d}{dy} \left( \left. \dfrac{\phi[F^{-1}(y)]}{ \left\vert\dfrac{dF(u)}{du}\right\vert_{u=F^{-1}(y)} }  \right) \right\vert_{y=0}  \right\} , \label{delta_int_disc_box_2} 
\end{eqnarray}
where the definition of the delta function in equation \eqref{DiracDelta_alternative_def} has been used. Here, $F^{-1}(y)$ represents the inverse function of $F(u)$.  

Performing the derivative in the equation \eqref{delta_int_disc_box_2}, and going back to the original variable $u$, the integral $I$ becomes,
\begin{eqnarray}
    I &=& \pi \Bigg\{ \delta[F(b)] \dfrac{\phi(b)}{F^\prime(b)} - \delta[F(a)] \dfrac{\phi(a)}{F^\prime(a)} + \dfrac{1}{\left\vert\dfrac{dF(u)}{du}\right\vert^3} \Big[ \phi(u_0) F^{\prime \prime}(u_0) \nonumber \\
    && -\phi^\prime(u_0) F^\prime(u_0) \Big] \Bigg\}.  \label{delta_int_disc_box_3} 
\end{eqnarray}

Thus, from the definition of the integral $I$ given in the equation \eqref{delta_int_disc_box}, and the result of the equation \eqref{delta_int_disc_box_3}, the following identity is obtained,
\begin{eqnarray}
    \lim_{\epsilon \rightarrow 0} \int_a^b du\, \phi(u) \, \dfrac{d}{dF}\left( \dfrac{\epsilon}{F^2+\epsilon^2} \right) &=& \pi \Bigg\{ \delta[F(b)] \dfrac{\phi(b)}{F^\prime(b)} - \delta[F(a)] \dfrac{\phi(a)}{F^\prime(a)} \nonumber \\
    && \hspace{-80pt} + \dfrac{1}{\left\vert\dfrac{dF(u_0)}{du}\right\vert^3} \Big[ \phi(u_0) F^{\prime \prime}(u_0) -\phi^\prime(u_0) F^\prime(u_0) \Big] \Bigg\}, \label{Derivative_delta_identity}
\end{eqnarray}
with the condition that $u_0 \in [a,b]$, being the only root of $F(u)$ in this interval. If $u_0 \notin [a,b]$, the right hand side of the equation \eqref{Derivative_delta_identity} vanishes.

\begin{figure}
    \centering
    \includegraphics[width=0.35\linewidth]{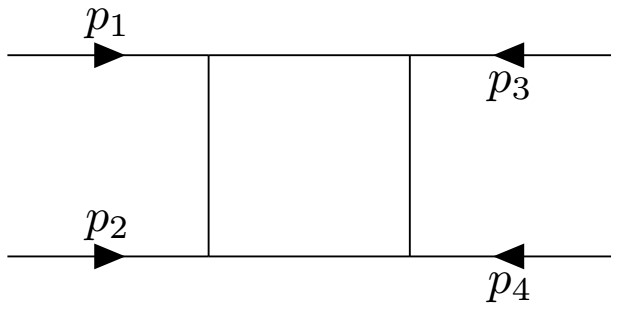}
    \caption{Feynman diagram for the box integral.}
    \label{Box_diagram}
\end{figure}

\section{Discontinuity of the massless box integral from the coproduct of the multiple polylogarithms} \label{Disc_Box_Append}

Figure \ref{Box_diagram} depicts the Feynman diagram for the box one-loop integral. In the onshell massless case, with one internal mass, the box integral is proportional to the function $D(s,t)$ \cite{Davydychev4p},
\begin{eqnarray}
    D(s,t) &=& \dfrac{1}{st\beta_{\hs\hT}}\Bigg( 2 \ln^2\left(\dfrac{\beta_{\hs\hT}+\beta_{\hs}}{\beta_{\hs\hT}+\beta_{\hT}}\right) + \ln\left(\dfrac{\beta_{\hs\hT}-\beta_{\hs}}{\beta_{\hs\hT}+\beta_{\hs}}\right) \ln\left(\dfrac{\beta_{\hs\hT}-\beta_{\hT}}{\beta_{\hs\hT}+\beta_{\hT}}\right) - \dfrac{\pi^2}{2} \nonumber \\
    && + \sum_{i=\hs,\hT}\left[ 2 \Li\left(\dfrac{\beta_i-1}{\beta_{\hs\hT}+\beta_i}\right) - 2 \Li\left(\dfrac{\beta_i-\beta_{\hs\hT}}{\beta_i+1} \right) - \ln^2\left(\dfrac{\beta_i+1}{\beta_{\hs\hT}+\beta_i}\right) \right] \Bigg), \nonumber \\ \label{Dav_fun}
\end{eqnarray}
where $\hs \equiv \frac{s}{4m^2}$, $\hT \equiv \frac{t}{4m^2}$, $\hu \equiv \frac{u}{4m^2}$, with $s$, $t$ and $u$ being the Mandelstam variables defined in the equation \eqref{Mandelstam}, and 
\begin{eqnarray}
    \beta_{x} = \sqrt{1-\dfrac{1}{x}}, \qquad \beta_{x y} = \sqrt{1-\dfrac{1}{x}-\dfrac{1}{y}}. \label{def_beta_functions}
\end{eqnarray}

Instead of using the equation \eqref{def_Disc}, the discontinuity of the function $D(s,t)$ across the branch cut with respect to $\bar{s}=\frac{s}{m^2}$ can be evaluated from the coproduct $\Delta$ of the multiple polylogarithms. As described in Refs \cite{DiscFeynmanInt} and \cite{Algebra_MPL}, if $\overline{\mathcal{H}}$ represents the $\mathbb{Q}$-vector space spanned by all multiple polylogaritms defined in equation \eqref{def_Gs},  then $\mathcal{H} = \overline{\mathcal{H}}/(\pi \overline{\mathcal{H}} )$, then $\mathcal{H}$ is a Hopf algebra equipped with the coproduct $\Delta: \mathcal{H} \rightarrow \mathcal{H} \otimes \mathcal{H}$. 

The coproduct is coassociative,
\begin{eqnarray}
    (\text{id}\otimes\Delta) \Delta = (\Delta \otimes \text{id})\Delta, 
\end{eqnarray}
and respects multiplication,
\begin{eqnarray}
    \Delta(a \cdot b) = \Delta(a) \cdot \Delta(b).
\end{eqnarray}

For logarithms and polylogarithms, the coproducts of these functions correspond to \cite{Algebra_MPL},
\begin{eqnarray}
    \Delta (\ln z) &=& 1 \otimes \ln z + \ln z \otimes 1, \nonumber \\
    \Delta (\text{Li}_n(z)) &=& 1 \otimes \text{Li}_n(z) + \text{Li}_n(z) \otimes 1 + \sum_{k=1}^{n-1} \text{Li}_{n-1}(z) \otimes \dfrac{\ln^k z}{k!}. \label{coproduct_log_and_polylog}
\end{eqnarray}

Additionally, the coproduct is defined such that,
\begin{eqnarray}
    \Delta(\pi) = \pi \otimes 1. 
\end{eqnarray}

Note that the coproduct in equation \eqref{coproduct_log_and_polylog} respects the weight of the logarithms and polylogarithms,\footnote{According to the definition of the multiple polylogarithms in equation \eqref{def_Gs}, the weigh of $\log z$ is equal to one, while the weight of $\text{Li}_n(z)$ is equal to $n$.} since the total weight of each term in the right hand side of equation \eqref{coproduct_log_and_polylog} (assuming that the weight of 1 is zer) match the weight of the functions $\ln(z)$ and $\text{Li}_n(z)$ in the left hand side. In general, this corresponds to an additional property of the coproduct. If $\mathcal{H}_n$ represents the subset of $\mathcal{H}$ with weigth $n$, then \cite{Algebra_MPL}
\begin{eqnarray}
    \mathcal{H}_n \xrightarrow{\Delta} \bigoplus_{k=0}^{n} \mathcal{H}_k \otimes \mathcal{H}_{n-k}.
\end{eqnarray}
Thus, the action of the coproduct on $\mathcal{H}_n$ can be decomposed as,
\begin{eqnarray}
    \Delta = \sum_{p+q=n} \Delta_{p,q},
\end{eqnarray}
where $\Delta_{p,q}$ is the part of the coproduct that takes values in $\mathcal{H}_p \otimes \mathcal{H}_q$. For example, according to the equation \eqref{coproduct_log_and_polylog},
\begin{eqnarray}
    \Delta_{1,1}( \Li(z) ) = - \ln(1-z) \otimes \ln z. \label{Delta11_dilog} 
\end{eqnarray}

As another example to be used later, consider the product of two logarithms, $\ln x \ln y$. Using the properties of the coproduct under multiplication and applying the equation \eqref{coproduct_log_and_polylog}, it can be shown that,
\begin{eqnarray}
    \Delta ( \ln x \ln y ) &=& \Delta (\ln x) \Delta(\ln y) \nonumber \\
    &=& 1 \otimes (\ln x \ln y) + (\ln x \ln y) \otimes 1 + \ln x \otimes \ln y + \ln y \otimes \ln x.
\end{eqnarray}
Thus,
\begin{eqnarray}
    \Delta_{1,1} ( \ln x \ln y ) = \ln x \otimes \ln y + \ln y \otimes \ln x. \label{Delta11_logs_prod}
\end{eqnarray}

Now, according to Ref. \cite{DiscFeynmanInt}, for an element $f_n$ of weight $n$ in $\overline{\mathcal{H}}$, its discontinuity $\text{Disc} f_n$ across a branch cut can be computed from the coproduct as,
\begin{eqnarray} 
    \dsc f_n \cong \mu \left[ (\dsc \otimes \text{id}) \Delta_{1,n-1} f_{n} \right], \label{disc_coproduct}
\end{eqnarray}
where $\cong$ denotes equivalence modulo $\pi^2$ and $\mu: \overline{\mathcal{H}} \rightarrow \overline{\mathcal{H}} \otimes \overline{\mathcal{H}} $ denotes the multiplication in $\overline{\mathcal{H}}$ which corresponds to multiplying the two factors of the coproduct. 

By applying the identity \eqref{disc_coproduct}, the discontinuity $\dsc_{\bar{s}} D(s,t)$ with the function $D(s,t)$ defined in equation \eqref{Dav_fun}, reads as
\begin{eqnarray}
    \dsc_{\bar{s}} D(s,t) = \mu [ (\dsc_{\bar{s}} \otimes \text{id}) \Delta_{1,1} D(s,t)  ], \label{Disc_s_Dst_v1}
\end{eqnarray}
where, from equations \eqref{Delta11_dilog} and \eqref{Delta11_logs_prod}, the expression for $\Delta_{1,1} D(s,t)$ corresponds to,
\begin{eqnarray}
    \Delta_{1,1} D(s,t) &=& \dfrac{1}{st\beta_{\hs\hT}} \Bigg[ \ln\left(\dfrac{\beta_{\hs\hT}-\beta_{\hs}}{\beta_{\hs\hT}+\beta_{\hs}}\right) \otimes \ln\left(\dfrac{\beta_{\hs\hT}-\beta_{\hT}}{\beta_{\hs\hT}+\beta_{\hT}}\right) \nonumber \\
    && \hspace{-30pt} + \ln\left(\dfrac{\beta_{\hs\hT}-\beta_{\hT}}{\beta_{\hs\hT}+\beta_{\hT}}\right) \otimes \ln\left(\dfrac{\beta_{\hs\hT}-\beta_{\hs}}{\beta_{\hs\hT}+\beta_{\hs}}\right) - 2 \sum_{i=\hs,\hT} \Big[  \ln\left(\dfrac{\beta_{\hs\hT}+1}{\beta_{\hs\hT}+\beta_i}\right)\otimes\ln\left(\dfrac{\beta_i-1}{\beta_{\hs\hT}+\beta_i}\right) \nonumber \\
    && \hspace{-30pt} - \ln\left(\dfrac{\beta_{\hs\hT}+1}{\beta_i+1}\right) \otimes \ln\left( -\dfrac{\beta_{\hs\hT}-\beta_{i}}{\beta_i+1} \right) \nonumber + \ln\left(\dfrac{\beta_{i}+1}{\beta_{\hs\hT}+\beta_i}\right)\otimes\ln\left(\dfrac{\beta_{i}+1}{\beta_{\hs\hT}+\beta_i}\right)  \Big] \nonumber \\
    && \hspace{-30pt} + 4 \ln\left( \dfrac{\beta_{\hs\hT}+\beta_{\hs}}{\beta_{\hs\hT}+\beta_{\hT}}  \right) \otimes \ln\left( \dfrac{\beta_{\hs\hT}+\beta_{\hs}}{\beta_{\hs\hT}+\beta_{\hT}} \right)\Bigg]. \nonumber \\ \label{Delta11_DavFun_st}
\end{eqnarray}
If the kinematical region $\bar{s}<-1<\bar{t}<0$ with $\bar{u}>4$ is considered, the equation \eqref{Disc_s_Dst_v1} reduces to,
\begin{eqnarray}
    \dsc_{\bar{s}} D(s,t) &=&\dfrac{1}{st\beta_{\hs\hT}} \Bigg\{ \left[ \dsc_{\bar{s}} \ln\left(\dfrac{\beta_{\hs\hT}-\beta_{\hs}}{\beta_{\hs\hT}+\beta_{\hs}}\right)  \right] \ln\left(\dfrac{\beta_{\hs\hT}-\beta_{\hT}}{\beta_{\hs\hT}+\beta_{\hT}}\right) \nonumber \\
    && + \left[\dsc_{\bar{s}} \ln\left(\dfrac{\beta_{\hs\hT}-\beta_{\hT}}{\beta_{\hs\hT}+\beta_{\hT}}\right) \right] \ln\left(\dfrac{\beta_{\hs\hT}-\beta_{\hs}}{\beta_{\hs\hT}+\beta_{\hs}}\right) \Bigg\}, \label{Disc_s_Dst_v2}
\end{eqnarray}

The logarithms of equation \eqref{Delta11_DavFun_st} that are not present in equation \eqref{Disc_s_Dst_v2} clearly do not present any discontinuity with respect to $\bar{s}$ in the kinematical region chosen previously. 

To evaluate the remaining discontinuities of equation \eqref{Disc_s_Dst_v2}, note that,
\begin{eqnarray}
    \beta_{\hs\hT}^2-\beta_{\hs}^2 &=& - \dfrac{4}{\bar{s}}, \nonumber \\
    \beta_{\hs\hT}^2-\beta_{\hT}^2 &=& - \dfrac{4}{\bar{t}}. \label{identityBetas}
\end{eqnarray}
Thus, equation \eqref{Disc_s_Dst_v2} can be rewritten as,
\begin{eqnarray}
    \dsc_{\bar{s}} D(s,t) &=& \dfrac{1}{st\beta_{\hs\hT}} \Bigg\{ \left[ \dsc_{\bar{s}} \ln\left(\dfrac{-\frac{4}{\bar{s}}}{(\beta_{\hs\hT}+\beta_{\hs})^2}\right)  \right] \ln\left(\dfrac{\beta_{\hs\hT}-\beta_{\hT}}{\beta_{\hs\hT}+\beta_{\hT}}\right) \nonumber \\
    && + \left[\dsc_{\bar{s}} \ln\left(\dfrac{-\frac{4}{\bar{t}}}{(\beta_{\hs\hT}+\beta_{\hT})^2}\right) \right] \ln\left(\dfrac{\beta_{\hs\hT}-\beta_{\hs}}{\beta_{\hs\hT}+\beta_{\hs}}\right)\Bigg\}, \label{Disc_s_Dst_v3}
\end{eqnarray}
which simplifies to 
\begin{eqnarray}
    \dsc_{\bar{s}} D(s,t) = 0, \label{Disc_s_Dst}
\end{eqnarray}
since in the kinematical region of interest the arguments of the logarithms in the square brackets are positive. 

Similarly, from the coproduct procedure, the discontinuity $\dsc_{\bar{s}} D(s,u)$ in the same kinematical region as before reads as, 
\begin{eqnarray}
    \dsc_{\bar{s}} D(s,u) &=&  \dfrac{1}{su\beta_{\hs\hu}}\left[ \dsc_{\bar{s}} \ln\left(-\dfrac{4}{\bar{u}}\right) \right] \ln\left(\dfrac{\beta_{\hs\hu}-\beta_{\hu}}{\beta_{\hs\hu}+\beta_{\hu}}\right) \nonumber \\
    &=& - \dfrac{1}{su\beta_{\hs\hu}} \left[ \dsc_{\bar{s}} \ln\left(-\bar{u}\right) \right] \ln\left(\dfrac{\beta_{\hs\hu}-\beta_{\hu}}{\beta_{\hs\hu}+\beta_{\hu}}\right),
\end{eqnarray}
which reduces to
\begin{eqnarray}
    \dsc_{\bar{s}} D(s,u) &=& - \dfrac{2\pi i}{su\beta_{\hs\hu}} \ln\left(\dfrac{\beta_{\hs\hu}-\beta_{\hu}}{\beta_{\hs\hu}+\beta_{\hu}}\right). \label{Disc_s_Dsu}
\end{eqnarray}

Finally, the computations used to derive the equation \eqref{Disc_s_Dsu}, lead to the following result,
\begin{eqnarray}
    \dsc_{\bar{s}} D(t,u) &=& - \dfrac{2\pi i}{tu\beta_{\hT\hu}} \ln\left(\dfrac{\beta_{\hT\hu}-\beta_{\hu}}{\beta_{\hT\hu}+\beta_{\hu}}\right). \label{Disc_s_Dtu}
\end{eqnarray}

\section{Two-point worldline integral} \label{Two_point_wl_int}

If in equation \eqref{scalar_phi3_masterformula} the value of the external particles is fixed to $N=2$, the expression for $\Gamma_{\text{1PI}}(p_1,p_2)$ corresponds to a bubble Feynman diagram, and will be proportional to,
\begin{eqnarray}
    B = \int_0^\infty \dfrac{dT}{T} \e^{-m^2T} T^{-\frac{D}{2}} \int_0^1 \int_0^1 d\tau_1 d\tau_2 \e^{p_1 \cdot p_2 G_{12}}. \label{Bubble_init_def}
\end{eqnarray}

After rescaling the worldline parameters $\tau_i$ as $\tau_i=T u_i$, and performing the integration on $T$ in $D=4-2\epsilon$ dimensions, the expression for the function $B$ in equation \eqref{Bubble_init_def} becomes,
\begin{eqnarray}
    B(p_1^2) = \dfrac{\mu^{2\epsilon}\Gamma(\epsilon)}{m^{2\epsilon}} \int_0^1 \int_0^1  \dfrac{1}{\left(1+\frac{p_1^2}{m^2}G_{12}\right)^\epsilon} du_1 du_2, \label{Bubble_afterT_int}
\end{eqnarray}
where it has been necessary to introduce the renormalization mass scale parameter $\mu$, since the bubble integrals are ultraviolet divergent in $D=4$ dimensions. Here, $\Gamma(\epsilon)$ denotes the gamma function, and the momentum conservation identity $p_2=-p_1$ has been used.

Setting $u_2=0$ in the equation \eqref{Bubble_afterT_int}, and expanding around $\epsilon$, the function $B(p_1^2)$ can be rewritten as,
\begin{eqnarray}
    B(p_1^2) = \dfrac{1}{\epsilon} - \gamma_E - \ln\left(\dfrac{m^2}{\mu^2}\right)-\int_0^1 du_1 \ln\left[1+\frac{p_1^2}{m^2}u_1(1-u_1)\right] + \mathcal{O}(\epsilon), \label{Bubble_intu1}
\end{eqnarray}
where $\gamma_E$ is the Euler's constant. 

Finally, after integrating on $u_1$, the expression for $B(p_1^2)$ becomes
\begin{eqnarray}
    B(p_1^2) = \dfrac{1}{\epsilon} - \gamma_E - \ln\left(\dfrac{m^2}{\mu^2}\right) + 2 + \overline{B}\left(-p_1^2\right) + \mathcal{O}(\epsilon), \label{Bubble_expression}
\end{eqnarray}
where the function $\overline{B}(q)$ is defined as,
\begin{eqnarray}
    \overline{B}(q) = \beta_{\hat{q}} \ln\left(\dfrac{\beta_{\hat{q}}-1}{\beta_{\hat{q}}+1}\right), \label{Bubblebar}
\end{eqnarray}
with $\hat{q}=q/(4m^2)$ and $\beta_x$ being defined in equation \eqref{def_beta_functions}.


\printbibliography

\end{document}